\documentclass{lmcs} 
\pdfoutput=1

\usepackage{lastpage}
\lmcsdoi{18}{3}{34}
\lmcsheading{}{\pageref{LastPage}}{}{}%
{Oct.~15,~2021}{Sep.~15,~2022}{}


\usepackage{etoolbox}
\usepackage{trfrac}
\usepackage{stmaryrd}
\usepackage{xspace}
\usepackage{alltt}
\usepackage{hyperref}
\usepackage[utf8]{inputenc}
\definecolor{icolor}{rgb}{0.15,0.75,0.2}
\definecolor{itcolor}{rgb}{0.15, 0.65, 0.6}
\definecolor{litcolor}{rgb}{0.7, 0.8, 0.8}
\definecolor{dcolor}{rgb}{1,0.6,0.2}
\definecolor{dtcolor}{rgb}{0.95, 0.45, 0.15}
\definecolor{myorange}{rgb}{1,0.55,0.0}
\definecolor{mypink}{rgb}{1.0,0.5,0.7}
\definecolor{myblue}{rgb}{0.45,0.65,0.9}
\definecolor{mypurple}{rgb}{0.6, 0.16, 0.6}
\definecolor{mylightgray}{rgb}{0.9, 0.9, 0.9}

\definecolor{agdablue}{rgb}{0.01, 0.1, 0.8}
\definecolor{agdagreen}{rgb}{0.1, 0.5, 0.2}
\definecolor{agdaorange}{rgb}{0.8, 0.2, 0.1}
\definecolor{agdared}{rgb}{0.6, 0.1, 0.1}

\newcommand{\black}[1]{\textcolor{black}{#1}}

\newcommand{\func}[1]{\textcolor{agdablue}{#1}}
\newcommand{\constr}[1]{\textcolor{agdagreen}{#1}}
\newcommand{\keyword}[1]{\textcolor{agdaorange}{#1}}

\newcommand{\Set}{\func{Set}}
\newcommand{\Ty}{\func{Ty}}
\newcommand{\CTy}{\func{CTy}}
\newcommand{\Tr}{\func{Tr}}
\newcommand{\Val}[2]{\func{Val[ \black{#1} ] \black{#2}}}
\newcommand{\CVal}[2]{\func{CVal[ \black{#1} ] \black{#2}}}
\newcommand{\Exp}[6]{\func{Exp[ \black{#1} ] \black{#2} \(\langle\) \black{#3} \(\rangle\) \black{#4} \(\langle\) \black{#5} \(\rangle\) \black{#6}}}
\newcommand{\idconttypefunc}{\func{id-cont-type}}
\newcommand{\Reduce}{\func{Reduce}\xspace}
\newcommand{\cpsv}{\func{cpsv}}
\newcommand{\cpse}{\func{cpse}}
\newcommand{\cpstau}[1]{\func{\(\llbracket\) \black{#1} \(\rrbracket\tau\)}}
\newcommand{\cpsmu}[1]{\func{\(\llbracket\) \black{#1} \(\rrbracket\mu\)}}
\newcommand{\bcirc}{\func{\(\circ\)}}
\newcommand{\Nat}{\constr{Nat}}
\newcommand{\Bool}{\constr{Bool}}
\newcommand{\Str}{\constr{Str}}
\newcommand{\Arr}[6]{\constr{\black{#1} \(\Rightarrow\) \black{#2} \(\langle\) \black{#3} \(\rangle\) \black{#4} \(\langle\) \black{#5} \(\rangle\) \black{#6}}}
\newcommand{\Bul}{\constr{\(\bullet\)}}
\newcommand{\Ext}[3]{\constr{\black{#1} \(\Rightarrow\!\langle\) \black{#2} \(\rangle\) \black{#3}}}
\newcommand{\Var}{\constr{Var}}
\newcommand{\Valcon}{\constr{Val}}
\newcommand{\App}{\constr{App}}
\newcommand{\Prompt}{\constr{Prompt}}
\newcommand{\RPrompt}{\constr{RPrompt}}
\newcommand{\darr}{\constr{\(\Rightarrow\)}}
\newcommand{\data}{\keyword{data}}
\newcommand{\where}{\keyword{where}}
\newcommand{\arr}{\(\rightarrow\)}
\newcommand{\tauty}{\(\tau\)}
\newcommand{\tauonety}{\(\tau_1\)}

\newcommand{\tautwoty}{\(\tau_2\)}
\newcommand{\alphaty}{\(\alpha\)}
\newcommand{\betaty}{\(\beta\)}
\newcommand{\betaprty}{\(\betapr\)}
\newcommand{\gammaty}{\(\gamma\)}

\newcommand{\deltaty}{\(\delta\)}

\newcommand{\mualphaty}{\(\mu\alpha\)}
\newcommand{\mubetaty}{\(\mu\beta\)}
\newcommand{\mugammaty}{\(\mu\gamma\)}
\newcommand{\mudeltaty}{\(\mu\delta\)}
\newcommand{\muity}{\(\mu_i\)}

\newcommand{\bdummy}{\!\!\func{\_}\!\!}
\newcommand{\gdummy}{\!\!\constr{\_}\!\!}

\newcommand{\lambdaF}{$\lambda_\mathcal{F}$\xspace}
\newcommand{\lambdaC}{$\lambda_C$\xspace}

\newcommand{\mtenv}{\bullet}
\newcommand{\extenv}[3]{#1, \, #2 : #3}

\newcommand{\wraptrail}[1]{\, \langle #1 \rangle \,}

\newcommand{\intty}{\texttt{int}\xspace}
\newcommand{\boolty}{\texttt{bool}\xspace}
\newcommand{\strty}{\texttt{string}\xspace}
\newcommand{\arrowty}[6]{#1 \rightarrow #2 \wraptrail{#3} #4 \wraptrail{#5} #6}
\newcommand{\arrowtyky}[5]{#1 \rightarrow #2, #3, #4 / #5}
\newcommand{\arrowtytwo}[3]{#1 \rightarrow #2 \rightarrow #3}
\newcommand{\arrowtypure}[2]{#1 \rightarrow #2}
\newcommand{\listty}[1]{\texttt{list}(#1)}

\newcommand{\taupr}{\tau^\prime}

\newcommand{\tauonepr}{{\tau_1}^\prime}

\newcommand{\tauthreepr}{{\tau_3}^\prime}

\newcommand{\betapr}{\beta^\prime}
\newcommand{\gammapr}{\gamma^\prime}

\newcommand{\abs}[2]{\lambda #1.\, #2}

\newcommand{\app}[2]{#1 \ #2}

\newcommand{\iszero}[1]{\texttt{is0} \ #1}
\newcommand{\iszerott}{\texttt{is0}\xspace}
\newcommand{\btos}[1]{\texttt{b2s} \ #1}
\newcommand{\btostt}{\texttt{b2s}\xspace}
\newcommand{\seq}[2]{#1; #2}

\newcommand{\shifttt}{\texttt{shift}\xspace}
\newcommand{\shiftsym}{\mathcal{S}\xspace}
\newcommand{\shift}[2]{\shiftsym #1.\, #2}
\newcommand{\shiftp}[2]{(\shift{#1}{#2})}
\newcommand{\resettt}{\texttt{reset}\xspace}

\newcommand{\reset}[1]{\langle #1 \rangle}

\newcommand{\shiftztt}{\texttt{shift0}\xspace}

\newcommand{\resetztt}{\texttt{reset0}\xspace}

\newcommand{\controltt}{\texttt{control}\xspace}

\newcommand{\controlsym}{\mathcal{F}\xspace}
\newcommand{\control}[2]{\controlsym #1.\, #2}
\newcommand{\controlp}[2]{(\control{#1}{#2})}
\newcommand{\prompttt}{\texttt{prompt}\xspace}
\newcommand{\promptsym}{\mathcal{P}\xspace}

\newcommand{\prompt}[1]{\langle #1 \rangle}

\newcommand{\controlztt}{\texttt{control0}\xspace}

\newcommand{\promptztt}{\texttt{prompt0}\xspace}

\newcommand{\mttrailty}{\bullet}
\newcommand{\exttrailty}[3]{#1 \rightarrow \wraptrail{#2} \, #3}

\newcommand{\mualpha}{\mu_{\alpha}}

\newcommand{\mubeta}{\mu_{\beta}}

\newcommand{\mugamma}{\mu_{\gamma}}
\newcommand{\mudelta}{\mu_{\delta}}

\newcommand{\idconttype}[3]{\mathsf{id\mathchar`-cont\mathchar`-type}(#1, #2, #3)}
\newcommand{\idconttypesf}{\textsf{id-cont-type}\xspace}
\newcommand{\compatible}[3]{\mathsf{compatible}(#1, #2, #3)}
\newcommand{\compatiblesf}{\textsf{compatible}\xspace}

\newcommand{\trail}[1]{\texttt{Trail}(#1)}
\newcommand{\recty}[2]{\mu #1.\, #2}

\newcommand{\cpsansty}[5]{(#1 \rightarrow #2 \rightarrow #3) \rightarrow #4 \rightarrow #5}
\newcommand{\cpsarrowty}[6]{#1 \rightarrow \cpsansty{#2}{#3}{#4}{#5}{#6}}
\newcommand{\cpscxtty}[3]{#1 \rightarrow #2 \rightarrow #3}

\newcommand{\abstwo}[3]{\abs{#1}{\abs{#2}{#3}}}
\newcommand{\absthree}[4]{\abs{#1}{\abs{#2}{\abs{#3}{#4}}}}

\newcommand{\apptwo}[3]{\app{\app{#1}{#2}}{#3}}
\newcommand{\appthree}[4]{\app{\app{\app{#1}{#2}}{#3}}{#4}}

\newcommand{\plus}[2]{#1 + #2}
\newcommand{\true}{\texttt{true}}
\newcommand{\false}{\texttt{false}}

\newcommand{\cons}[2]{#1 :: #2}
\newcommand{\consop}{::}
\newcommand{\consuscore}{\_\!::\!\_}
\newcommand{\append}[2]{#1 \, @ \, #2}
\newcommand{\appendop}{@}
\newcommand{\appenduscore}{\_@\_}
\newcommand{\casett}{\texttt{case}}
\newcommand{\oftt}{\texttt{of}}

\newcommand{\caseof}[5]{\casett \ #1 \ \oftt \ #2 \Rightarrow #3
\mid #4 \Rightarrow #5}

\newcommand{\epr}{e^\prime}
\newcommand{\eonepr}{{e_1}^\prime}
\newcommand{\etwopr}{{e_2}^\prime}
\newcommand{\tpr}{t_1}

\newcommand{\kpr}{k_1}

\newcommand{\Fpr}{F^\prime}

\newcommand{\idcont}{k_{id}}

\newcommand{\mttrail}{()}

\newcommand{\cps}[1]{\llbracket #1 \rrbracket}

\newcommand{\cpsty}[1]{#1^\ast}
\newcommand{\trans}{\leadsto}
\newcommand{\transp}{\leadsto_p}

\newcommand{\subst}[3]{#1 \, [#3 / #2]}

\newcommand{\reduce}{\leadsto}
\newcommand{\reduces}{\leadsto^\ast}
\newcommand{\hole}{[]}
\newcommand{\holedot}{[.]}

\newcommand{\judge}[7]{#1 \vdash #2 : #3 \wraptrail{#4} #5 \wraptrail{#6} #7}
\newcommand{\judgep}[5]{\judge{#1}{#2}{#3}{#4}{#5}{#4}{#5}}

\newcommand{\judgeky}[6]{#1 \vdash #2 : #3, #4, #5 / #6}

\newcommand{\judgec}[3]{#1 \vdash #2 : #3}

\newcommand{\judgetrans}[8]{\judge{#1}{#2}{#3}{#4}{#5}{#6}{#7} \ \shade{\trans #8}}
\newcommand{\judgeptrans}[4]{\judgep{#1}{#2}{#3} \ \shade{\transp #4}}

\newcommand{\rulename}[1]{(\textsc{#1})}
\newcommand{\typing}[3]{\trfrac{#1}{#2} \ \rulename{#3}}

\newcommand{\ie}{\textit{i.e.}\xspace}

\newcommand{\ssep}{\hspace{3mm}}
\newcommand{\lsep}{\hspace{1cm}}

\newcommand{\leftbf}[1]{\begin{flushleft} \textbf{#1} \end{flushleft}}

\newcommand{\shade}[1]{\colorbox{mylightgray}{$#1$}}
\newcommand{\im}[1]{\vspace{3mm} \begin{center} $#1$ \end{center} \vspace{3mm}}

\newcommand{\para}[1]{\vspace{3mm} \noindent \textbf{#1.}\xspace}

\newtheoremstyle{break}{}{}{\itshape}{}{\bfseries}{.}{\newline}{}
\theoremstyle{break}

\newtheoremstyle{case}{}{}{}{}{}{}{\newline}{}
\theoremstyle{case}
\newtheorem{case}{Case}

\AtBeginEnvironment{proof}{\setcounter{case}{0}
                           \setcounter{subcase}{0}
                           \setcounter{subsubcase}{0}}
\AtBeginEnvironment{case}{\setcounter{subcase}{0}
                          \setcounter{subsubcase}{0}}
\AtBeginEnvironment{subcase}{\setcounter{subsubcase}{0}}

\begin{document}

\title[A Functional Abstraction of Typed Invocation Contexts]{A Functional Abstraction of \\ Typed Invocation Contexts}

\thanks{The authors are grateful to the anonymous reviewers for their 
thoughtful comments, which improved this paper in various ways.
This work was supported in part by JSPS KAKENHI under Grant No.~JP18H03218, 
No.~JP19K24339, and No.~JP22H03563.}

\author[Y.~Cong]{Youyou Cong\lmcsorcid{0000-0003-2315-6182}}[a]

\author[C.~Ishio]{Chiaki Ishio}[b]

\author[K.~Honda]{Kaho Honda}[b]

\author[K.~Asai]{Kenichi Asai}[b]

\address{Tokyo Institute of Technology, 2-12-1, Ookayama, Meguro-ku, Tokyo, Japan}
\email{cong@c.titech.ac.jp}

\address{Ochanomizu University, 2-1-1, Otsuka, Bunkyo-ku, Tokyo, Japan}
\email{\{ishio.chiaki, honda.kaho, asai\}@is.ocha.ac.jp}

\begin{abstract}
In their paper ``A Functional Abstraction of Typed Contexts'', Danvy and
Filinski show how to derive a monomorphic type system of the \shifttt and
\resettt operators from a CPS semantics.
In this paper, we show how this method scales to Felleisen's \controltt
and \prompttt operators.
Compared to \shifttt and \resettt, \controltt and \prompttt exhibit a
more dynamic behavior, in that they can manipulate a \emph{trail} of
contexts surrounding the invocation of previously captured continuations.
Our key observation is that, by adopting a functional representation of
trails in the CPS semantics, we can derive a type system that encodes
all and only constraints imposed by the CPS semantics.

\end{abstract}

\maketitle

\section{Introduction}
\label{sec:intro}

Delimited continuations have been proven useful in diverse domains.
Their applications range from representation of monadic
effects~\cite{filinski-representing}, to formalization of partial
evaluation~\cite{danvy-tdpe}, and to implementation of automatic
differentiation~\cite{wang-demystifying}.
As a means to handle delimited continuations, researchers have designed
a variety of \emph{control operators}~\cite{felleisen-prompt,
danvy-abstracting, gunter-cupto, dyvbig-monadic, materzok-subtyping}.
Among them, Danvy and Filinski's~\cite{danvy-abstracting} \shifttt/\resettt
operators have a solid theoretical foundation:
there exist a simple CPS translation to the pure
$\lambda$-calculus~\cite{danvy-abstracting}, an expressive type system that
allows modification of answer types~\cite{danvy-context}, and a set of
equational axioms that are sound and complete with respect to the CPS
translation~\cite{kameyama-axiom}.
A variation of \shifttt/\resettt, known as \shiftztt/\resetztt, also have
similar theoretical artifacts, due to recent work by Materzok and
Biernacki~\cite{materzok-subtyping, materzok-axiomatizing}.
Other variants, however, are not as well-understood as the aforementioned
ones, because there is currently no simple CPS semantics for those variants.

Understanding the subtleties of control operators is important, especially
given the rapid adoption of \emph{algebraic effects and
handlers}~\cite{plotkin-handler, bauer-tutorial} observed in the past decade.
Effect handlers can be thought of as a generalization of exception handlers
that provide access to the continuation surrounding an exception.
It is known that effect handlers have a close connection with control 
operators~\cite{forster-macro, pirog-typedeq}, and they are often implemented 
using control operators provided by the host 
language~\cite{kammar-handler, kiselyov-eff}.
This means a well-established theory of control operators is crucial for
safer and more efficient implementation of effect handlers.

In this paper, we formalize a typed calculus of \controltt and \prompttt,
a pair of control operators proposed by Felleisen~\cite{felleisen-prompt}.
These operators bring an interesting behavior into programs: when a
captured continuation $k$ is invoked, the subsequent computation may
capture the context surrounding the invocation of $k$.
From a practical point of view, the ability to manipulate invocation
contexts is useful for implementing sophisticated algorithms, such as list
reversing~\cite{biernacki-brics} and breadth-first
traversal~\cite{biernacki-bft}.
From a theoretical perspective, on the other hand, this ability makes it
hard to type programs in a way that fully reflects their runtime behavior.

We address the challenge with typing by rigorously following Danvy and
Fillinski's~\cite{danvy-context} recipe for building a type system of a
delimited control calculus.
The idea is to analyze the denotational semantics of the calculus ---
essentially a CPS translation --- and identify all the constraints that are
necessary for making the denotation well-typed.
In fact, the recipe has been applied to the \controltt and \prompttt
operators before~\cite{kameyama-dynamic}, but the type system obtained
is not satisfactory for two reasons.
First, the type system imposes certain restrictions on the contexts in
which a captured continuation may be invoked.
Second, the type system does not precisely describe the way contexts compose
and propagate during evaluation.
We show that, by choosing a right representation of invocation contexts
in the CPS translation, we can build a type system without such limitations.

Below is a summary of our specific contributions:

\begin{itemize}
\item We present a type system of \controltt and \prompttt that accounts
  for manipulation of invocation contexts.
  The type system is the \controltt/\prompttt-equivalent of Danvy and
  Filinski's~\cite{danvy-context} type system for \shifttt/\resettt, in
  that it incorporates all and only constraints that are imposed by the
  monomorphically typed CPS translation.
  
\item We prove two properties of our type system: type soundness and type 
  preservation of the CPS translation.
  We also provide a partial proof of termination, which does not hold for the 
  existing type system of \controltt and \prompttt~\cite{kameyama-dynamic}.

\item We further design a more fine-grained type system that distinguishes
  between pure and impure expressions, and define a selective CPS
  translation based on the type system.
  The selective translation is crucial for practical implementation of
  \controltt and \prompttt, as it generates less administrative redexes.

\item We formalize our calculi and CPS translations in the Agda proof
  assistant~\cite{norell-phd}.
  The formalization serves as the mechanized proofs of the theorems stated
  in the paper (except the progress theorem), and is available online at:

  \begin{center}
  \url{https://github.com/YouyouCong/lmcs-artifact}
  \end{center}
\end{itemize}

We begin with an informal account of \controltt and \prompttt
(Section~\ref{sec:control}), highlighting the dynamic behavior of these
operators.
We next formalize an untyped calculus of \controltt/\prompttt
(Section~\ref{sec:lambdaf}) and its CPS translation (Section~\ref{sec:cps}).
Then, from the CPS translation, we derive a type system of our calculus
(Section~\ref{sec:type}), and prove its properties (Section~\ref{sec:prop}).
Having established a basic typing principle, we refine the calculus with
the notion of purity, and define a selective CPS translation
(Section~\ref{sec:selective}).
After that, we briefly explain how we developed our Agda formalization.
Lastly, we discuss related work (Section~\ref{sec:related}) and conclude
with future directions (Section~\ref{sec:conclusion}).

\paragraph{Relation to Prior Work}
This is an updated and extended version of our previous paper~\cite{cong-fscd}.
The primary new contribution of this version is the fine-grained calculus and
selective CPS translation developed in Section~\ref{sec:selective}.
We also used the extra space to describe our Agda formalization and expand
various discussions.

\section{Control and Prompt}
\label{sec:control}

In this section, we familiarize the reader with \controltt and \prompttt
by walking through two examples.
These examples exhibit the dynamic behavior of \controltt and \prompttt,
as well as the heterogeneous nature of captured contexts.

\subsection{The Dynamic Behavior of Control}
\label{sec:control:dynamic}

The \controltt and \prompttt operators are control primitives for capturing
and delimiting continuations.
To see how these operators behave, let us look at the following program:

\vspace{-2mm}

\begin{align}
\prompt{\controlp{k_1}{2 * (\app{k_1}{5})} +
        \controlp{k_2}{3 + (\app{k_2}{8})}} + 13 \tag{1}
\end{align}

\vspace{2mm}

\noindent Throughout the paper, we write $\controlsym$ to mean \controltt and
$\prompt{}$ to mean \prompttt.
Under the call-by-value, left-to-right evaluation strategy, the above
program evaluates in the following way:

\vspace{-2mm}

\begin{align*}
& \prompt{
    \controlp{k_1}{2 * (\app{k_1}{5})} +
    \controlp{k_2}{3 + (\app{k_2}{8})}} + 13 \\
&= \prompt{
     \subst{2 * (\app{k_1}{5})}{k_1}
           {\abs{x}{x} + \controlp{k_2}{3 + (\app{k_2}{8})}}
   } + 13 \\
&= \prompt{2 * (5 + \controlp{k_2}{3 + (\app{k_2}{8})})} + 13 \\
&= \prompt{
     \subst{3 + (\app{k_2}{8})}{k_2}{\abs{x}{2 * (5 + x)}}
   } + 13 \\
&= \prompt{3 + (2 * (5 + 8))} + 13 \\
&= 42
\end{align*}

\vspace{2mm}

\noindent The first \controltt operator captures the delimited context up
to the enclosing \prompttt, namely
$\holedot + \controlp{k_2}{3 + (\app{k_2}{8})}$ (where $\holedot$ denotes
a hole).
The captured context is then reified into a function
$\abs{x}{x + \controlp{k_2}{3 + (\app{k_2}{8})}}$, and evaluation
shifts to the body $2 * (\app{k_1}{5})$, where $k_1$ is the reified
continuation.
After $\beta$-reducing the invocation of $k_1$, we obtain another
\controltt in the evaluation position.
This \controltt captures the context $2 * (5 + \holedot)$, which
is a composition of \emph{two} contexts: the addition context originally
surrounding the \controltt construct, and the multiplication context
surrounding the invocation of $k_1$.
The context is then reified into a function
$\abs{x}{2 * (5 + x)}$, and evaluation shifts to the body
$3 + (\app{k_2}{8})$, where $k_2$ is the reified continuation.
By $\beta$-reducing the invocation of $k_2$, we obtain the expression
$3 + (2 * (5 + 8))$, where the original delimited context, the
invocation context of $k_1$, and the invocation context of $k_2$ are all
composed together.
The expression returns the value $29$ to the enclosing \prompttt clause,
and the evaluation of the whole program finishes with $42$.

From this example, we observe that a \controltt operator can capture the
context surrounding the invocation of a previously captured continuation.
More generally, \controltt may capture a \emph{trail} of such invocation
contexts.
The ability comes from the absence of the delimiter in the body of
captured continuations.
Indeed, if we replace \controltt with \shifttt ($\shiftsym$) in the above
program, the second \shifttt would have no access to the context
$2 * \holedot$, since the first \shifttt would insert a \resettt into
the continuation $k_1$.
As a consequence, the program yields a different result $45$.

\vspace{-2mm}

\begin{align*}
& \reset{
    \shiftp{k_1}{2 * (\app{k_1}{5})} +
    \shiftp{k_2}{3 + (\app{k_2}{8})}} + 13 \\
&= \reset{
     \subst{2 * (\app{k_1}{5})}{k_1}
           {\abs{x}{\reset{x + \shiftp{k_2}{3 + (\app{k_2}{8})}}}}
   } + 13 \\
&= \reset{2 * \reset{5 + \shiftp{k_2}{3 + (\app{k_2}{8})}}} + 13 \\
&= \reset{
     2 * \reset{\subst{3 + (\app{k_2}{8})}{k_2}{\abs{x}{\reset{5 + x}}}}
   } + 13 \\
&= \reset{2 * \reset{3 + \reset{5 + 8}}} + 13 \\
&= 45
\end{align*}

\vspace{2mm}

In the continuations literature, \controltt/\prompttt are classified as
\emph{dynamic}, because they require us to actually run a program in
order to see what continuation a \controltt captures.
In contrast, \shifttt/\resettt are called \emph{static}, as they allow us
to determine the extent of continuations from the lexical information of
a program.

\subsection{The Heterogeneous Nature of Trails}
\label{sec:control:hetero}

In the example discussed above, we only had contexts whose input and
output types are both \intty.
In practice, however, we may wish to deal with contexts having different
input and output types.
Here is such an example:

\vspace{-2mm}

\begin{align}
\prompt{\controlp{k_1}{\iszero{(\app{k_1}{5})}} +
        \controlp{k_2}{\btos{(\app{k_2}{8})}}} \tag{2}
\end{align}

\vspace{2mm}

\noindent We assume two primitive functions: \iszerott, which tells us
if a given integer is zero or not, and \btostt, which converts a boolean
into a string \texttt{"\true"} or \texttt{"\false"}.
Below is the reduction sequence of the above program:

\vspace{-2mm}

\begin{align*}
& \prompt{(\control{k_1}{\iszero{(\app{k_1}{5})}}) +
          (\control{k_2}{\btos{(\app{k_2}{8})}})} \\
& = \prompt{\subst{\iszero{(\app{k_1}{5})}}
                  {k_1}
                  {\abs{x}{x + (\control{k_2}{\btos{(\app{k_2}{8})}})}}} \\
& = \prompt{\iszero{(5 + (\control{k_2}{\btos{(\app{k_2}{8})}}))}} \\
& = \prompt{\subst{\btos{(\app{k_2}{8})}}{k_2}{\abs{x}{\iszero{(5 + x)}}}} \\
& = \prompt{\btos{(\iszero{(5 + 8)})}} \\
& = \prompt{\btos{(\iszero{13})}} \\
& = \prompt{\btos{\false}} \\
& = \prompt{\texttt{"\false"}} \\
& = \texttt{"\false"}
\end{align*}

\vspace{2mm}

\noindent The first \controltt operator captures the delimited context
$\holedot + \controlp{k_2}{\btos{(\app{k_2}{8})}}$,
and invokes it in the context $\iszero{\holedot}$.
The second \controltt captures the context
$\iszero{(\plus{5}{\holedot})}$,
and invokes it in the context $\btos{\holedot}$.
By composing these contexts, and by plugging $8$ into the hole, we obtain
$\btos{(\iszero{(5 + 8)})}$, which evaluates to the string \texttt{"$\false$"}.

From this example, we observe that a trail of invocation contexts can be
\emph{heterogeneous}.
In particular, the input and output types of each context can be different.
Of course, the invocation contexts in a heterogeneous trail must align
in a way that makes their composition type-safe.
The type safety holds when the input type of an invocation context is equal
to the output type of the previous context.
It is easy to see that example (2) meets this condition: the original
delimited context ($5 + \holedot$) is \intty-to-\intty, the first invocation
context ($\iszero{\holedot}$) is \intty-to-\boolty, and the second invocation
context ($\btos{\holedot}$) is \boolty-to-\strty.
Therefore, the application $\btos{(\app{k_2}{8})}$ is well-typed.

It turns out that example (2) is judged ill-typed by the existing type
system for \controltt and \prompttt~\cite{kameyama-dynamic}.
This is because the type system imposes the following restrictions on the
type of invocation contexts.

\begin{itemize}
\item All invocation contexts within a \prompttt clause must have the
  same type.
\item For each invocation context, the input and output types must be
  the same.
\end{itemize}

We claim that, a fully expressive type system of \controltt and \prompttt
should be more flexible about the type of invocation contexts.
Now the question is: Is it possible to allow such flexibility?
Our answer is ``yes''.
As we will see in Section~\ref{sec:type}, we can build a type system
that accommodates invocation contexts having different types, and that
accepts example (2) as a well-typed program.

\newcommand{\LambdaF}{
\begin{figure}

\leftbf{Syntax}
\begin{align*}
v &::= n \mid x \mid \abs{x}{e} & \text{Values} \\
e &::= v \mid \app{e}{e} \mid \control{k}{e} \mid \prompt{e}
  & \text{Expressions}
\end{align*}

\vspace{4mm}

\leftbf{Evaluation Contexts}
\begin{align*}
E &::= \hole \mid \app{E}{e} \mid \app{v}{E} \mid
       \prompt{E} & \text{General Contexts} \\
F &::= \hole \mid \app{F}{e} \mid \app{v}{F}
       & \text{Pure Contexts}
\end{align*}

\vspace{4mm}

\leftbf{Reduction Rules}
\begin{align*}
E[\app{(\abs{x}{e})}{v}] &\reduce E[\subst{e}{x}{v}] & \rulename{$\beta$} \\
E[\prompt{F[\control{k}{e}]}] &\reduce E[\prompt{\subst{e}{k}{\abs{x}{F[x]}}}]
  & \rulename{$\controlsym$} \\
E[\prompt{v}] &\reduce E[v] & \rulename{$\promptsym$}
\end{align*}

\caption{Syntax and Reduction of \lambdaF}
\label{fig:lambdaf}
\end{figure}
}

\section{\lambdaF: A Calculus of Control and Prompt}
\label{sec:lambdaf}

\LambdaF

Having seen how \controltt and \prompttt behave, we define a formal
calculus of these operators.
In Figure~\ref{fig:lambdaf}, we present \lambdaF, a (currently untyped)
$\lambda$-calculus extended with \controltt and \prompttt.
The calculus has a separate syntactic category for values, which includes
integers, variables, and abstractions.
Expressions consist of values, application, and delimited control
constructs \controltt and \prompttt.

We equip \lambdaF with a call-by-value, left-to-right evaluation strategy.
As is usual with delimited control calculi, there are two groups of
evaluation contexts: general contexts ($E$) and pure contexts ($F$).
Their difference is that general contexts may contain \prompttt surrounding
a hole, while pure contexts can never have such \prompttt.
The distinction is used in the reduction rule \rulename{$\controlsym$} of
\controltt, which says, \controltt always captures the context up to the
nearest enclosing \prompttt.
In the reduct, we see that the body of a captured continuation is \emph{not}
surrounded by \prompttt, as we observed in the previous section.
On the other hand, the body of \controltt is evaluated in a \prompttt
clause.
The reduction rule \rulename{$\promptsym$} for \prompttt simply removes a
delimiter surrounding a value.

\newcommand{\LambdaC}{
\begin{figure}

\leftbf{Syntax}
\vspace{-1mm}
\begin{align*}
v, t &::= n \mid x \mid \abs{x}{e} \mid \mttrail & \text{Values} \\
e    &::= v \mid \app{e}{e} \mid (\caseof{t}{\mttrail}{e}{k}{e})
     & \text{Expressions}
\end{align*}

\vspace{4mm}

\leftbf{Evaluation Contexts}
\begin{align*}
E &::= \hole \mid \app{E}{e} \mid \app{v}{E} \mid (\caseof{E}{\mttrail}{e}{k}{e})
\end{align*}

\vspace{4mm}

\leftbf{Reduction Rules}
\begin{align*}
E[\app{(\abs{x}{e})}{v}] &\reduce E[\subst{e}{x}{v}]
  & \rulename{$\beta$} \\
E[\caseof{\mttrail}{\mttrail}{e_1}{k}{e_2}] &\reduce E[e_1]
  & \rulename{case-$\mttrail$} \\
E[\caseof{v}{\mttrail}{e_1}{k}{e_2}] &\reduce E[\subst{e_2}{k}{v}]
  \quad \text{where } v \neq \mttrail
  & \rulename{case-$k$}
\end{align*}

\caption{Syntax and Reduction of \lambdaC}
\label{fig:lambdac}
\end{figure}
}

\newcommand{\CPS}{
\begin{figure}

\begin{align*}
\cps{n} &=
  \abstwo{k}{t}{\apptwo{k}{n}{t}} \\
\cps{x} &=
  \abstwo{k}{t}{\apptwo{k}{x}{t}} \\
\cps{\abs{x}{e}} &=
  \abstwo{k}{t}
         {\apptwo{k}
                 {(\absthree{x}{\kpr}{\tpr}
                            {\apptwo{\cps{e}}{\kpr}{\tpr}})}{t}} \\
\cps{\app{e_1}{e_2}} &=
  \abstwo{k}{t}
         {\apptwo{\cps{e_1}}
                 {(\abstwo{v_1}{t_1}
                          {\apptwo{\cps{e_2}}
                                  {(\abstwo{v_2}{t_2}
                                           {\appthree{v_1}{v_2}{k}{t_2}})}
                                  {t_1}})}
                 {t}} \\
\cps{\control{c}{e}} &=
  \abstwo{k}{t}
         {\apptwo{\subst{\cps{e}}
                        {c}
                        {\absthree{x}{\kpr}{\tpr}
                                  {\apptwo{k}{x}
                                          {(\append{t}{(\cons{\kpr}{\tpr})})}}}}
                 {\idcont}
                 {\mttrail}} \\
\cps{\prompt{e}} &=
  \abstwo{k}{t}
         {\apptwo{k}{(\apptwo{\cps{e}}{\idcont}{\mttrail})}{t}} \\
& \\
\idcont &= \abstwo{v}{t}{\caseof{t}{\mttrail}{v}{k}{\apptwo{k}{v}{\mttrail}}} \\
\appenduscore &=
  \abstwo{t}{\tpr}
         {\caseof{t}{\mttrail}{\tpr}{k}{\cons{k}{\tpr}}} \\
\consuscore &=
  \abstwo{k}{t}
         {\caseof{t}
                 {\mttrail}{k}
                 {\kpr}{\abstwo{v}{\tpr}
                               {\apptwo{k}{v}{(\cons{\kpr}{\tpr})}}}}
\end{align*}

\caption{CPS Translation of \lambdaF Expressions}
\label{fig:cps}
\end{figure}
}

\section{CPS Translation}
\label{sec:cps}

In the previous section, we defined the semantics of \controltt and
\prompttt directly on \lambdaF expressions.
Another way to define the semantics is to convert \lambdaF expressions into
the plain $\lambda$-calculus, via a continuation-passing style
(CPS)~\cite{danvy-context} translation.
In this section, we give a CPS translation of \controltt and \prompttt.

When the source calculus has \controltt and \prompttt, a CPS translation
exposes continuations and trails of invocation contexts.
Among these, trails can be represented either as a list of
functions~\cite{biernacki-brics, biernacki-toplas} or as a composition of
functions~\cite{shan-simulation}.
Previous work~\cite{kameyama-dynamic} on typing \controltt and \prompttt
adopts the list representation, because it is conceptually simpler.
We, however, adopt the functional representation, as it makes it easier
to build a fully expressive type system (see Section~\ref{sec:type}).

\subsection{\lambdaC: Target Calculus of CPS Translation}
\label{sec:cps:lambdac}

\LambdaC

In Figure~\ref{fig:lambdac}, we define the target calculus of the CPS
translation, which we call \lambdaC.
The calculus is a pure, call-by-value $\lambda$-calculus featuring the
unit value $\mttrail$, which represents an empty trail, and a case
analysis construct, which allows inspection of trails.
Note that a non-empty trail is represented as a regular function.
Note also that we use $v$ and $t$ as both meta-level and object-level
variables denoting values and trails.

As in \lambdaF, we evaluate \lambdaC programs under a call-by-value,
left-to-right strategy.
The choice of evaluation strategy in the target calculus is not important
in our setting, but it matters if the target calculus has computational
effects (such as non-termination and I/O), because the CPS image of
\controltt and \prompttt has non-tail calls.

\subsection{The CPS Translation}
\label{sec:cps:trans}

\CPS

In Figure~\ref{fig:cps}, we present the CPS translation $\cps{\_}$ from
\lambdaF to \lambdaC, which is equivalent to the translation given by
Shan~\cite{shan-simulation}.
The translation converts an expression into a function that takes in a
continuation $k$ and a trail $t$.
The continuation receives a value and a trail and produces an answer.
The trail is the composition of the invocation contexts encountered
so far.
Here is an informal typing of continuations and trails:

\vspace{-2mm}

\begin{align*}
\mathtt{Cont} &= \arrowtytwo{\mathtt{Val}}{\mathtt{Trail}}{\mathtt{Val}} \\ 
\mathtt{Trail} &= 1 + \mathtt{Cont}
\end{align*}

\vspace{2mm}

\noindent The formal typing is obtained by specifying the type of 
each component (see Section~\ref{sec:type:trail}).

Below, we detail the translation of three representative constructs:
variables, \prompttt, and \controltt.

\para{Variables}
The translation of a variable is an $\eta$-expanded version of
the standard, call-by-value translation.
The trivial use of the current trail $t$ communicates the fact that a
variable can never change the trail during evaluation.
More generally, the CPS translation of a pure expression uniformly
calls the continuation with an unmodified trail.

\para{Prompt}
The translation of \prompttt has the same structure as the translation
of variables, because \prompttt forms a pure expression.
The translated body $\cps{e}$ is run with the identity continuation
$\idcont$ and an empty trail $\mttrail$\footnote{The identity continuation
$\idcont$ and the empty trail $\mttrail$ correspond to the \texttt{send}
function and the \texttt{\#f} value of Shan~\cite{shan-simulation},
respectively.},
describing how \prompttt resets the context of evaluation.
Note that, in this CPS translation, the identity continuation is \emph{not}
the identity function.
It receives a value $v$ and a trail $t$, and behaves differently depending on
whether $t$ is empty or not.
When $t$ is empty, the identity continuation simply returns $v$.
When $t$ is non-empty, $t$ must be a function composed of one or more
invocation contexts, which looks like $\abs{x}{E_n[... \, E_1[x] \, ...]}$.
In this case, the identity continuation builds an expression
$E_n[... \, E_1[v] \, ...]$ by calling the trail with $v$ and $\mttrail$.

\para{Control}
The translation of \controltt shares the same pattern with the
translation of \prompttt, because its body is evaluated in a \prompttt
clause (as defined by the \rulename{$\controlsym$} rule in
Figure~\ref{fig:lambdaf}).
The translated body $\cps{e}$ is applied a substitution that replaces the
variable $c$ with the trail $\append{t}{(\cons{\kpr}{\tpr})}$, describing
how an invocation of a captured continuation extends the trail\footnote{
There is a superficial difference between our CPS translation and
Shan's original translation~\cite{shan-simulation}.
In our translation rule for \controltt, the continuation variable $c$ is
replaced by the function
$\absthree{x}{\kpr}{\tpr}{\apptwo{k}{x}{(\append{t}{(\cons{\kpr}{\tpr})})}}$.
In Shan's translation, $c$ is replaced by
$\absthree{x}{\kpr}{\tpr}{\apptwo{(\cons{k}{t})}{x}{(\cons{\kpr}{\tpr})}}$.
However, by expanding the definition of $\appendop$ and $\consop$, we can
easily see that the two functions are equivalent.
We prefer the one that uses $\appendop$ because it is closer to the
abstract machine given by Biernacki et al.~\cite{biernacki-toplas}, as well
as the list-based CPS translation derived from it.}.
The extension is necessary for correctly computing the continuations
captured during the invocation of $c$.
In particular, it makes the invocation context $\kpr$ of $c$ accessible to
the \controltt operators that are executed during the invocation of $c$.
Recall that, in this CPS translation, trails are represented as functions.
The $\appendop$ and $\consop$ operators are thus defined as a function
producing a function\footnote{The $\consop$ function is equivalent to
Shan's \texttt{compose} function.}.
More specifically, these operators compose contexts in a first-captured,
first-called manner (as we can see from the second clause of $\consop$).
Notice that $\consop$ is defined as a \emph{recursive} function\footnote{
While recursive, the $\consop$ function is guaranteed to terminate, as the
types of the two arguments become smaller in every three successive
recursive calls (or they reach the base case in fewer steps).}.
The reason is that, when extending a trail $t$ with a continuation $k$, we
need to produce a function that takes in a trail $\tpr$, which in turn must
be composed with a continuation $\kpr$.

The CPS translation is correct with respect to the definitional abstract
machine given by Biernacka et al.~\cite{biernacka-operational}.
The statement is proved by Shan~\cite{shan-simulation}, using the functional
correspondence~\cite{ager-functional} between evaluators and abstract
machines.
We are also developing a mechanized proof of correctness in Agda.
The goal is to show that, if two \lambdaF expressions are related by
reduction, their CPS images are related by observational equivalence.
We have proved all cases but the \controltt reduction (rule
\rulename{$\controlsym$} in Figure~\ref{fig:lambdaf}), which requires
careful reasoning on the equivalence of trails.

As a last note, let us mention here that the alternative CPS translation
of \controltt and \prompttt, where trails are represented as lists, can be
obtained by replacing $\mttrail$ with the empty list, and the two
operations $\appendop$ and $\consop$ with ones that work on lists.

\newcommand{\Type}{
\begin{figure}

\leftbf{Syntax of Types and Typing Environments}
\begin{align*}
\tau, \alpha, \beta &::=
  \intty \mid \arrowty{\tau}{\tau}{\mualpha}{\alpha}{\mubeta}{\beta} &
  \text{Expression Types} \\
\mu, \mualpha, \mubeta &::= \mttrailty \mid \exttrailty{\tau}{\mu}{\tau} &
  \text{Trail Types} \\
\Gamma &::=
  \mtenv \mid \extenv{\Gamma}{x}{\tau} &
  \text{Typing Environments}
\end{align*}

\vspace{5mm}

\leftbf{Typing Rules}
\[
\typing{}
       {\judgep{\Gamma}{n}{\intty}{\mualpha}{\alpha}}
       {Int}
\lsep
\typing{x : \tau \in \Gamma}
       {\judgep{\Gamma}{x}{\tau}{\mualpha}{\alpha}}
       {Var}
\]

\vspace{4mm}

\[
\typing{\judge{\extenv{\Gamma}{x}{\tau_1}}
              {e}{\tau_2}{\mualpha}{\alpha}{\mubeta}{\beta}}
       {\judgep{\Gamma}{\abs{x}{e}}
               {(\arrowty{\tau_1}{\tau_2}{\mualpha}{\alpha}{\mubeta}{\beta})}
               {\mugamma}{\gamma}}
       {Abs}
\]

\vspace{4mm}

\[
\typing{\begin{trgather}
        \judge{\Gamma}{e_1}
              {(\arrowty{\tau_1}{\tau_2}{\mualpha}{\alpha}{\mubeta}{\beta})}
              {\mugamma}{\gamma}{\mudelta}{\delta} \\
        \judge{\Gamma}{e_2}{\tau_1}{\mubeta}{\beta}{\mugamma}{\gamma}
        \end{trgather}}
       {\judge{\Gamma}{\app{e_1}{e_2}}{\tau_2}
              {\mualpha}{\alpha}{\mudelta}{\delta}}
       {App}
\]

\vspace{4mm}

\[
\typing{\begin{trgather}
        \judge{\extenv{\Gamma}{k}
                      {\arrowty{\tau}{\tau_1}{\mu_1}{\tauonepr}{\mu_2}{\alpha}}}
              {e}{\gamma}{\mu_i}{\gammapr}{\mttrailty}{\beta} \\
        \idconttype{\gamma}{\mu_i}{\gammapr} \\
        \compatible{(\exttrailty{\tau_1}{\mu_1}{\tauonepr})}{\mu_2}{\mu_0} \\
        \compatible{\mubeta}{\mu_0}{\mualpha}
        \end{trgather}}
       {\judge{\Gamma}{\control{k}{e}}{\tau}{\mualpha}{\alpha}{\mubeta}{\beta}}
       {Control}
\]

\vspace{4mm}

\[
\typing{\begin{trgather}
        \judge{\Gamma}{e}{\beta}{\mu_i}{\betapr}{\mttrailty}{\tau} \\
        \idconttype{\beta}{\mu_i}{\betapr}
        \end{trgather}}
       {\judgep{\Gamma}{\prompt{e}}{\tau}{\mualpha}{\alpha}}
       {Prompt}
\]

\caption{Type System of \lambdaF}
\label{fig:type}
\end{figure}
}

\newcommand{\Aux}{
\begin{figure}

\begin{align*}
\idconttype{\tau}{\mttrailty}{\taupr} &= \tau \equiv \taupr \\
& \hspace{5mm} \text{(first branch of $\idcont$)} \\
\idconttype{\tau}{(\exttrailty{\tau_1}{\mu}{\tau_1^\prime})}{\taupr} &=
  (\tau \equiv \tau_1) \wedge
  (\taupr \equiv \tau_1^\prime) \wedge
  (\mu \equiv \mttrailty) \\
& \hspace{5mm} \text{(second branch of $\idcont$)} \\
& \\
\compatible{\mttrailty}{\mu_2}{\mu_3} &= \mu_2 \equiv \mu_3 \\
& \hspace{5mm} \text{(first branch of $\appendop$)} \\
\compatible{\exttrailty{\tau_1}{\mu_1}{\tau_1^\prime}}{\mttrailty}{\mu_3} &= \exttrailty{\tau_1}{\mu_1}{\tau_1^\prime} \equiv \mu_3 \\
& \hspace{5mm} \text{(first branch of $\consop$)} \\
\compatible{(\exttrailty{\tau_1}{\mu_1}{\tau_1^\prime})}
           {(\exttrailty{\tau_2}{\mu_2}{\tau_2^\prime})}
           {\mttrailty} &= \bot \\
& \hspace{5mm} \text{(no counterpart)} \\
\compatible{(\exttrailty{\tau_1}{\mu_1}{\tau_1^\prime})}
           {(\exttrailty{\tau_2}{\mu_2}{\tau_2^\prime})}
           {(\exttrailty{\tau_3}{\mu_3}{\tau_3^\prime})} &=
  (\tau_1 \equiv \tau_3) \wedge
  (\tau_1^\prime \equiv \tau_3^\prime) \, \wedge \\
  & \hspace{5mm} (\compatiblesf ((\exttrailty{\tau_2}{\mu_2}{\tau_2^\prime}), \\
  & \hspace{2.7cm} \mu_3, \mu_1)) \\
& \hspace{5mm} \text{(second branch of $\consop$)}
\end{align*}

\caption{Auxiliary Relations}
\label{fig:aux}
\end{figure}
}

\section{Type System}
\label{sec:type}

The CPS translation defined in the previous section shows how \controltt
and \prompttt manipulate continuations and trails.
Guided by this behavior, we now derive a type system of \lambdaF.
We proceed in three steps.
First, we specify the syntax of trail types (Section~\ref{sec:type:trail}).
Next, we identify an appropriate form of typing judgment
(Section~\ref{sec:type:judge}).
Lastly, we define the typing rules of individual syntactic constructs
(Section~\ref{sec:type:rule}).
In each step, we contrast our outcome with its counterpart in Kameyama
and Yonezawa's~\cite{kameyama-dynamic} type system, showing how different
representations of trails in the CPS translation lead to different typing
principles.

\subsection{Syntax of Trail Types}
\label{sec:type:trail}

Recall from Section~\ref{sec:cps:lambdac} that, in \lambdaC, trails have
two possible forms: $\mttrail$ or a function.
Correspondingly, in \lambdaF, trail types $\mu$ are defined by a
two-clause grammar: $\mttrailty \mid \exttrailty{\tau}{\mu}{\taupr}$.
The latter type is interpreted in the following way.

\begin{itemize}
\item The trail accepts a value of type $\tau$.
\item The trail is to be composed with a context of type $\mu$.
\item After the composition, the trail produces a value of type $\taupr$.
\end{itemize}

\noindent Put differently, $\tau$ is the input type of the innermost
invocation context, $\taupr$ is the output type of the context to be
composed in the future, and $\mu$ is the type of this future context.

To get some intuition about trail typing, let us revisit example (2)
from Section~\ref{sec:control:hetero}.

\vspace{-2mm}

\begin{align*}
& \prompt{\controlp{k_1}{\iszero{(\app{k_1}{5})}} +
          \controlp{k_2}{\btos{(\app{k_2}{8})}}} \\
& = \prompt{\subst{\iszero{(\app{k_1}{5})}}
                  {k_1}
                  {\abs{x}{x + \controlp{k_2}{\btos{(\app{k_2}{8})}}}}} \\
& = \prompt{\iszero{(5 + \controlp{k_2}{\btos{(\app{k_2}{8})}})}} \\
& = \prompt{\subst{\btos{(\app{k_2}{8})}}{k_2}{\abs{x}{\iszero{(5 + x)}}}} \\
& = \prompt{\btos{(\iszero{(5 + 8)})}} \\
& = \texttt{"\false"}
\end{align*}

\vspace{2mm}

When evaluation goes inside of \prompttt, the trail is initialized to an
empty one $\mttrail$.
Such a trail is naturally given type $\mttrailty$.

When the continuation $k_1$ is invoked, the trail is extended with the
context $\iszero{\holedot}$.
This context will be composed with the invocation context $\btos{\holedot}$ of
$k_2$ later in the reduction sequence.
Therefore, the trail at this point is given type
$\exttrailty{\intty}{\exttrailty{\boolty}{\mttrailty}{\strty}}{\strty}$,
consisting of the input type of \iszerott, the type of \btostt, and the
output type of \btostt.

When the continuation $k_2$ is invoked, the trail is extended with the
context $\btos{\holedot}$ (hence the whole trail looks like
$\btos{(\iszero{\holedot})}$).
This context will not be composed with any further contexts in the
subsequent steps of reduction.
Therefore, the trail at this point is given type
$\exttrailty{\intty}{\mttrailty}{\strty}$, consisting of the input type of
\iszerott, the type of an empty trail, and the output type of \btostt.

Observe that our trail types can be inhabited by heterogeneous trails
(e.g., $\btos{(\iszero{\holedot})}$), where the input and output types of
each invocation context are different.
The flexibility is exactly what we expect an expressive type system of 
\controltt and \prompttt to have, as we stated in 
Section~\ref{sec:control:hetero}.

\vspace{3mm}

\paragraph{Comparison with Previous Work}
In the CPS translation of Kameyama and Yonezawa \cite{kameyama-dynamic}, a
trail is treated as a list of invocation contexts.
Such a list is given a recursive type $\trail{\rho}$ defined as follows:

\im{
\trail{\rho} = \recty{X}{\listty{\arrowtytwo{\rho}{X}{\rho}}}
}

\noindent We can easily see that the definition restricts the type of
invocation contexts in two ways.
First, all invocation contexts in a trail must have the same type.
This is because lists are homogeneous by definition.
Second, each invocation context must have equal input and output types.
This is a direct consequence of the first restriction.
The two restrictions prevent one from invoking a continuation in a
context such as $\iszero{\holedot}$ or $\btos{\holedot}$.
Moreover, the use of the list type makes empty and non-empty trails
indistinguishable at the level of types, and extension of trails
undetectable in types.
On the other hand, the limited expressiveness allows one to use an ordinary
expression type (such as \intty, instead of a type designed specifically
for trails) to encode the information of trails in the
\controltt/\prompttt calculus.
That is, if a trail has type $\trail{\rho}$ in the CPS world, it has type
$\rho$ in the direct-style world.

\subsection{Typing Judgment}
\label{sec:type:judge}

Recall from Section~\ref{sec:cps:trans} that a \lambdaF expression $e$
is CPS-translated into a function $\abstwo{k}{t}{\epr}$.
Assuming $e$ has type $\tau$, the translated function can be typed in
the following way:

\im{
\cps{e^\tau} = \abstwo{k^{\cpscxtty{\tau}{\mualpha}{\alpha}}}{t^{\mubeta}}{{\epr}^\beta}
}

\noindent Here, $\alpha$ and $\beta$ are answer types, representing the
return type of the enclosing \prompttt before and after evaluation of $e$.
It is well-known that delimited control can make the two answer types
distinct~\cite{danvy-context}, and since answer types are needed to decide
typability of programs, they must be integrated into the typing
judgment.
The other pair of types, $\mubeta$ and $\mualpha$, are trail types,
representing the composition of invocation contexts encountered before
and after evaluation of $e$.
We have seen that an invocation of a captured continuation can change the
type of the trail, and since trail types are also relevant to typability
of programs, they have to be integrated into the typing judgment.

Summing up the above discussion, we conclude that a fully expressive typing
judgment for \controltt and \prompttt must carry five types, as follows:

\im{
\judge{\Gamma}{e}{\tau}{\mualpha}{\alpha}{\mubeta}{\beta}
}

\noindent We place the types in the same order as their appearance in the
annotated CPS expression.
That is, the first three types $\tau$, $\mualpha$, and $\alpha$ correspond
to the continuation of $e$, the next one $\mubeta$ represents the trail
required by $e$, and the last one $\beta$ stands for the eventual value
returned by $e$.
We will hereafter call $\alpha$ and $\beta$ initial and final answer types,
and $\mubeta$ and $\mualpha$ initial and final trail types --- be careful
of the direction in which answer types and trail types change.

With the typing judgment specified, we can define the syntax of expression
types of \lambdaF (Figure~\ref{fig:type}).
Expression types are formed with the integer type $\intty$ and the arrow type
$\arrowty{\tau_1}{\tau_2}{\mualpha}{\alpha}{\mubeta}{\beta}$.
Notice that the codomain of arrow types carries five components.
These types represent the control effect of a function's body, and correspond
exactly to the five types that appear in a typing judgment.

\vspace{3mm}

\paragraph{Comparison with Previous Work}
In the type system of Kameyama and Yonezawa~\cite{kameyama-dynamic},
a CPS-translated expression is typed in the following way:

\im{
\abstwo{k^{\cpscxtty{\tau}{\trail{\rho}}{\alpha}}}{t^{\trail{\rho}}}{{\epr}^\beta}
}

\noindent The typing is obviously not as flexible as ours, because the two
trail types are equal.
This constraint is imposed by the list representation of trails: even if an
expression extends the trail during evaluation, the type of the trail stays
the same.
Thus, Kameyama and Yonezawa arrive at a typing judgment carrying four types,
with the last one ($\rho$) representing the information of trails:

\im{
\judgeky{\Gamma}{e}{\tau}{\alpha}{\beta}{\rho}
}

\noindent Correspondingly, they assign direct-style functions an arrow
type of the form $\arrowtyky{\tau_1}{\tau_2}{\alpha}{\beta}{\rho}$.

\subsection{Typing Rules}
\label{sec:type:rule}

\Type

\Aux

Using the trail types and typing judgment from the previous subsections,
we define the typing rules of \lambdaF (Figure~\ref{fig:type}).
As in the previous section, we elaborate the typing rules of variables,
\prompttt, and \controltt.

\paragraph{Variables}
Recall that the CPS translation of variables is an $\eta$-expanded version
of the standard translation.
If we annotate the types of each subexpression, a translated variable would
look like:

\im{
\abstwo{k^{\cpscxtty{\tau}{\mualpha}{\alpha}}}{t^{\mualpha}}
       {(\apptwo{k}{x}{t})^{\alpha}}
}

\noindent We see duplicate occurrences of the answer type $\alpha$ and the
trail type $\mualpha$.
The duplication arises from the application $\apptwo{k}{x}{t}$, and reflects
the fact that a variable cannot change the answer type or the trail type.
By a straightforward conversion from the annotated expression into a typing
judgment, we obtain rule \rulename{Var} in Figure~\ref{fig:type}.
In general, when the subject of a typing judgment is a pure expression, the
answer types and trail types both coincide.

\paragraph{Prompt}
We next analyze the CPS translation of \prompttt, again with type annotations.

\im{
\abstwo{k^{\cpscxtty{\tau}{\mualpha}{\alpha}}}{t^{\mualpha}}
       {(\apptwo{k}{(\apptwo{\cps{e}^{\cpsansty{\beta}{\mu_i}{\betapr}{\mttrailty}{\tau}}}
                            {\idcont}{\mttrail})}{t})^{\alpha}}
}

\noindent As $\prompt{e}$ is a pure expression, we again have equal answer
types $\alpha$ and trail types $\mualpha$ for the whole expression.
The initial trail type $\mttrailty$ and final answer type $\tau$ of $e$ are
determined by the application $\apptwo{\cps{e}}{\idcont}{\mttrail}$ and
$\app{k}{(\apptwo{\cps{e}}{\idcont}{\mttrail})}$, respectively.
What is left is to ensure that the application of $\cps{e}$ to the identity
continuation $\idcont$ is type-safe.
In our type system, we use a relation $\idconttype{\tau}{\mu}{\taupr}$ to
ensure this type safety.
The relation holds when the type $\cpscxtty{\tau}{\mu}{\taupr}$ can be
assigned to the identity continuation.
The valid combination of $\tau$, $\mu$, and $\taupr$ is derived from the
definition of the identity continuation, repeated below.

\im{
\abstwo{v^{\tau}}{t^{\mu}}
       {\caseof{t}{\mttrail}{v^{\taupr}}
               {k}{(\apptwo{k}{v}{\mttrail})^{\taupr}}}
}

When $t$ is an empty trail $\mttrail$ of type $\mttrailty$, the return
value of $\idcont$ is $v$, which has type $\tau$.
Since the expected return type of $\idcont$ is $\taupr$, we need the
equality $\tau \equiv \taupr$.

When $t$ is a non-empty trail $k$ of type $\cpscxtty{\tau_1}{\mu}{\tauonepr}$,
the return value of $\idcont$ is the result of the application
$\apptwo{k}{v}{\mttrail}$, which has type $\tauonepr$.
Since the expected return type of $\idcont$ is $\taupr$, we need the
equality $\taupr \equiv \tauonepr$.
Furthermore, since $k$ must accept $v$ and $\mttrail$ as arguments, we need
the equalities $\tau \equiv \tau_1$ and $\mu \equiv \mttrailty$.

We define \idconttypesf as an encoding of these constraints, and in the rule
\rulename{Prompt}, we use the premise $\idconttype{\beta}{\mu_i}{\betapr}$ to
constrain the type of the continuation of $e$.
Now, it is statically guaranteed that $e$ will be safely evaluated in an
empty context.

\paragraph{Control}
Lastly, we apply the same method to \controltt.
Here is the annotated CPS translation:

\im{
\abstwo{k^{\cpscxtty{\tau}{\mualpha}{\alpha}}}{t^{\mubeta}}
       {\apptwo{\subst{\cps{e}^{\cpsansty{\gamma}{\mu_i}{\gammapr}{\mttrailty}{\beta}}}
                      {c}
                      {\absthree{x^\tau}{{\kpr}^{\cpscxtty{\tau_1}{\mu_1}{\tauonepr}}}{{\tpr}^{\mu_2}}
                               {\apptwo{k}{x}{(\append{t}{(\cons{\kpr}{\tpr})^{\mu_0}})}}}}
                {\idcont}
                {\mttrail}}
}

\noindent As the body $e$ of \controltt is evaluated in a \prompttt clause,
we again have an empty initial trail type for $e$, and we know that the types
$\gamma$, $\mu_i$, and $\gammapr$ must satisfy the \idconttypesf relation.
What is left is to ensure that the composition of contexts in
$\append{t}{(\cons{\kpr}{\tpr})}$ is type-safe.
In our type system, we use a relation $\compatible{\mu_1}{\mu_2}{\mu_3}$ to
ensure this type safety.
The relation holds when composing a context of type $\mu_1$ and another
context of type $\mu_2$ results in a context of type $\mu_3$.
Intuitively, the relation can be thought of as an addition over trail types
(or equivalently, the type-level counterpart of append and cons).
The valid combination of $\mu_1$, $\mu_2$, and $\mu_3$ is derived from the
definition of the $\appendop$ and $\consop$ functions.

\vspace{-4mm}

\begin{align*}
\append{t^{\mu_1}}{{\tpr}^{\mu_2}} &=
\caseof{t}{\mttrail}{{\tpr}^{\mu_3}}{k}{(\cons{k}{\tpr})^{\mu_3}} \\
\cons{k^{\cpscxtty{\tau_1}{\mu_1}{\tauonepr}}}{t^{\mu_2}} &=
  \caseof{t}{\mttrail}{k^{\cpscxtty{\tau_3}{\mu_3}{\tauthreepr}}}
         {\kpr}{(\abstwo{v}{\tpr}{\apptwo{k}{v}{(\cons{\kpr}{\tpr})}})^{\cpscxtty{\tau_3}{\mu_3}{\tauthreepr}}}
\end{align*}

The first clause of $\appendop$ and that of $\consop$ are straightforward:
they tell us that the empty trail type $\mttrailty$ serves as the left and
right identity of the addition.

The second clause of $\consop$ requires more careful reasoning.
The return value of this case is the result of the application
$\apptwo{k}{v}{(\cons{\kpr}{\tpr})}$, which has type $\tauonepr$.
Since the expected return type of $\consop$ is $\tauthreepr$,
we need the equality $\tauonepr \equiv \tauthreepr$.
Moreover, since $k$ must accept $v$ and $\cons{\kpr}{\tpr}$ as arguments,
we need the equality $\tau_1 \equiv \tau_3$, as well as the compatibility
of $\mu_2$ (type of $\kpr$), $\mu_3$ (type of $\tpr$), and $\mu_1$ (type
of the trail expected by $k$).


We define \compatiblesf as an encoding of these constraints, and in the
\rulename{Control} rule, we use two instances of this relation to
constrain the type of contexts appearing in $\append{t}{(\cons{\kpr}{\tpr})}$.
Among the two instances, the first one
$\compatible{(\exttrailty{\tau_1}{\mu_1}{\tauonepr})}{\mu_2}{\mu_0}$
states that consing $\kpr$ to $\tpr$ is type-safe, and the result has type
$\mu_0$.
The second one $\compatible{\mubeta}{\mu_0}{\mualpha}$ states that appending
$t$ to $\cons{\kpr}{\tpr}$ is type-safe, and the result has type $\mualpha$,
which is required by the continuation $k$ of the whole \controltt expression.

By designing \rulename{Control} and \rulename{Prompt} in this way, we have
achieved the goal of this work: a type system that encodes all and only
restrictions that arise from the CPS semantics of \controltt and \prompttt.
More precisely, we use the \idconttypesf and \compatiblesf relations to
enforce type-safe manipulation of contexts, while allowing the full
flexibility in the type of contexts being manipulated.

\paragraph{Comparison with Previous Work}
In the type system of Kameyama and Yonezawa \cite{kameyama-dynamic}, the
typing rules for \controltt and \prompttt are defined as follows:

\im{
\typing{\judgeky{\extenv{\Gamma}{k}{\arrowtyky{\tau}{\rho}{\rho}{\alpha}{\rho}}}
                {e}{\gamma}{\gamma}{\beta}{\gamma}}
       {\judgeky{\Gamma}{\control{k}{e}}{\tau}{\alpha}{\beta}{\rho}}
       {Control}
\lsep
\typing{\judgeky{\Gamma}{e}{\rho}{\rho}{\tau}{\rho}}
       {\judgeky{\Gamma}{\prompt{e}}{\tau}{\alpha}{\alpha}{\sigma}}
       {Prompt}
}

\noindent The rules are simpler than the corresponding rules in our type
system.
In particular, there is no equivalent of \idconttypesf or \compatiblesf, since
the homogeneous nature of trails makes those relations trivial.

A closer look at rule \rulename{Prompt} further reveals how \prompttt
determines the trail type of its body.
To spell out, the first and third occurrences of $\rho$ in the premise
mean that, if an expression $e$ is surrounded by a \prompttt whose body
is of type $\rho$, then the invocation contexts in the trail of $e$ all
have the same input / output type $\rho$.

\subsection{Typing Heterogeneous Trails}
\label{sec:type:hetero}

We now show that example (2) from Section~\ref{sec:control:hetero} is
judged well-typed in \lambdaF\footnote{An implementation \texttt{exp4} of
this example is included in \texttt{lambdaf.agda} in the artifact repository.}.
The well-typedness of the whole program largely relies on the well-typedness
of the two \controltt constructs, so let us look at the typing of these
constructs:

\im{
\judge{}{\control{k_1}{\iszero{(\app{k_1}{5})}}}{\intty}
      {\mu_1}{\strty}{\mttrailty}{\strty}
}

\vspace{-3mm}

\im{
\judge{}{\control{k_2}{\btos{(\app{k_2}{8})}}}{\intty}
      {\mu_2}{\strty}{\mu_1}{\strty}
}

\noindent For brevity, we write $\mu_1$ to mean
$\exttrailty{\intty}{\exttrailty{\boolty}{\mttrailty}{\strty}}{\strty}$,
and $\mu_2$ to mean $\exttrailty{\intty}{\mttrailty}{\strty}$.
We can see how the trail type changes from empty ($\mttrailty$), to one
that refers to a future context ($\mu_1$), and to one that mentions no
further context ($\mu_2$).
In particular, $\mu_2$ is the result of ``adding'' $\mu_1$ and the type
of $\btos{\holedot}$; that is, the invocation of $k_2$ \emph{discharges}
the future context awaited by $\iszero{\holedot}$.
The trail type $\mu_2$ serves as the final trail type of the body of the
enclosing \prompttt, and as it allows us to establish the \idconttypesf
relation required by \rulename{Prompt}, we can conclude that the whole
program is well-typed.

\newcommand{\TypeC}{
\begin{figure}

\leftbf{Syntax of Types}
\vspace{-2mm}
\begin{align*}
\tau &= \intty \mid \arrowtypure{\tau}{\tau} \mid \mttrailty
\end{align*}

\vspace{2mm}

\leftbf{Typing Rules}
\[
\typing{}
       {\judgec{\Gamma}{n}{\intty}}
       {Int}
\lsep
\typing{x : \tau \in \Gamma}
       {\judgec{\Gamma}{x}{\tau}}
       {Var}
\lsep
\typing{\judgec{\extenv{\Gamma}{x}{\tau_1}}{e}{\tau_2}}
       {\judgec{\Gamma}{\abs{x}{e}}
               {\arrowtypure{\tau_1}{\tau_2}}}
       {Abs}
\]

\vspace{2mm}

\[
\typing{}
       {\judgec{\Gamma}{\mttrail}{\mttrailty}}
       {Unit}
\lsep
\typing{\judgec{\Gamma}{e_1}{\arrowtypure{\tau_1}{\tau_2}} \ssep
        \judgec{\Gamma}{e_2}{\tau_1}}
       {\judgec{\Gamma}{\app{e_1}{e_2}}{\tau_2}}
       {App}
\]

\vspace{2mm}

\[
\typing{\begin{trgather}
        \judgec{\Gamma}{t}{\cpsty{\mu}} \\
        \judgec{\Gamma}{e_1}{\tau}
          \quad \text{assuming } \mu \equiv \mttrailty \\
        \judgec{\extenv{\Gamma}{k}{\cpsty{\mu}}}{e_2}{\tau}
          \quad \text{assuming } \mu \equiv \exttrailty{\tau_1}{\mu_1}{\tauonepr}
        \end{trgather}}
       {\judgec{\Gamma}{\caseof{t}{\mttrail}{e_1}{k}{e_2}}{\tau}}
       {Case}
\]

\caption{Type System of \lambdaC}
\label{fig:typec}
\end{figure}
}

\newcommand{\CPSType}{
\begin{figure}

\leftbf{Translation of Expression Types}
\vspace{-2mm}
\begin{align*}
\cpsty{\intty} &= \intty \\
\cpsty{(\arrowty{\tau_1}{\tau_2}{\mualpha}{\alpha}{\mubeta}{\beta})} &=
  \cpsarrowty{\cpsty{\tau_1}}
             {\cpsty{\tau_2}}{\cpsty{\mualpha}}{\cpsty{\alpha}}
             {\cpsty{\mubeta}}{\cpsty{\beta}}
\end{align*}

\vspace{4mm}

\leftbf{Translation of Trail Types}
\vspace{-2mm}
\begin{align*}
\cpsty{\mttrailty} &= \mttrailty \\
\cpsty{(\exttrailty{\tau}{\mu}{\taupr})} &=
  \cpscxtty{\cpsty{\tau}}{\cpsty{\mu}}{\cpsty{{\taupr}}}
\end{align*}

\caption{CPS Translation of \lambdaF Types}
\label{fig:cpstype}
\end{figure}
}

\section{Properties}
\label{sec:prop}

The type system of \lambdaF enjoys several pleasant properties.
In this section, we discuss two properties with a complete proof and one 
property with a partial proof.

\subsection{Type Soundness}
\label{sec:prop:sound}

First, the type system is sound, that is, well-typed programs do not go
wrong~\cite{milner-polymorphism}.
Following Wright and Felleisen~\cite{wright-syntactic}, we prove type
soundness via the preservation and progress theorems.

\begin{thm}[Preservation]
\label{thm:preservation}
If $\judge{\Gamma}{e}{\tau}{\mualpha}{\alpha}{\mubeta}{\beta}$
and $e \reduce \epr$,
then $\judge{\Gamma}{\epr}{\tau}{\mualpha}{\alpha}{\mubeta}{\beta}$.
\end{thm}

\begin{proof}
The proof is by induction on the typing derivation\footnote{The proof is
implemented as the \texttt{Reduce} relation in \texttt{lambdaf-red.agda}.}.
Note that, to prove this property, we need to define a set of typing rules
for evaluation contexts.
\end{proof}

\begin{thm}[Progress]
\label{thm:progress}
If $\judge{\mtenv}{e}{\tau}{\mualpha}{\alpha}{\mubeta}{\beta}$,
then either (i) $e$ is a value, (ii) $e$ takes a step, or
(iii) $e$ is a stuck term of the form $F[\control{k}{\epr}]$.
\end{thm}

\begin{proof}
The proof is by induction on the typing derivation\footnote{The progress
property is proven by hand, not by Agda, due to the difficulty of
expressing closed expressions while managing bindings via parametric
higher-order abstract syntax~\cite{chlipala-phoas}.
We provide the proof in Appendix~\ref{app:progress}.}.
The third alternative is commonly found in the progress property of
effectful calculi~\cite{asai-polymorphic, xie-evidently}, meaning the
presence of unhandled effects.
\end{proof}

\begin{thm}[Type Soundness]
\label{thm:soundness}
If $\judge{\mtenv}{e}{\tau}{\mttrailty}{\alpha}{\mttrailty}{\alpha}$,
then evaluation of $e$ does not get stuck.
\end{thm}

\begin{proof}
The statement is a direct implication of preservation and progress.
Note that the empty final trail type means $e$ does not have control
effects.
This is justified by the fact that the final trail type $\mualpha$ in
the typing rule \rulename{Control} must be non-empty, as the trail is
composed of a non-empty trail of type $\exttrailty{\tau_1}{\mu_1}{\tauonepr}$.
\end{proof}

\TypeC

\CPSType

\subsection{Type Preservation of CPS Translation}
\label{sec:prop:preservation}

Secondly, our CPS translation preserves typing, \ie, it converts a
well-typed \lambdaF expression into a well-typed \lambdaC expression.
To establish this theorem, we define the type system of \lambdaC
(Figure~\ref{fig:typec}) and the CPS translation $\cpsty{}$ on \lambdaF
types (Figure~\ref{fig:cpstype}).

Let us elaborate on rule \rulename{Case} in Figure~\ref{fig:typec},
which is the only non-trivial typing rule.
This rule is used to type the case analysis construct in the three
auxiliary functions of the CPS translation, namely $\idcont$, $\appendop$,
and $\consop$.
A key feature of this rule is that it uses meta-level equality assumptions
$\mu \equiv \mttrailty$ and $\mu \equiv \exttrailty{\tau_1}{\mu_1}{\tauonepr}$
to type the two branches\footnote{The use of equality assumptions in
\rulename{Case} is inspired by
\emph{dependent pattern matching}~\cite{coquand-pattern} available in
dependently typed languages.
Our case analysis is weaker than the dependent variant, in that the
return type only depends on the \emph{type} of the scrutinee, not on the
scrutinee itself.
For the formal definition, we refer the reader to our Agda implementation 
(\texttt{lambdac.agda}).}.
These assumptions, together with the \idconttypesf and \compatiblesf relations,
fill in the gap between the expected and actual return types.
As an illustration, let us elaborate the typing of $\idcont$:

\im{
\abstwo{v^{\tau}}{t^{\mu}}
       {\caseof{t}{\mttrail}{v^{\taupr}}
               {k}{(\apptwo{k}{v}{\mttrail})^{\taupr}}}
}

\noindent In the first branch, we see an inconsistency between the
expected return type $\taupr$ and the actual return type $\tau$.
However, by the typing rules defined in Figure~\ref{fig:type}, we know
that $\idcont$ is used only when the relation $\idconttype{\tau}{\mu}{\taupr}$
holds.
Moreover, by the definition of \idconttypesf, if $\mu \equiv \mttrailty$, we
have $\tau \equiv \taupr$.
The equality assumption $\mu \equiv \mttrailty$ made available by rule
\rulename{Case} allows us to derive $\tau \equiv \taupr$ and conclude
that the first branch has the correct type.
Similarly, in the second branch, we use the equality assumption
$\mu \equiv \exttrailty{\tau_1}{\mu_1}{\tau_1^\prime}$ to
derive $\tau \equiv \tau_1$, $\taupr \equiv \tau_1^\prime$, and
$\mu_1 \equiv \mttrailty$, which imply the well-typedness of the
application $\apptwo{k}{v}{\mttrail}$.
The $\appendop$ and $\consop$ functions can be typed in an analogous
way.

Now we state the type preservation theorem.

\begin{thm}[Type Preservation of CPS Translation]
\label{thm:preservationcps}
If $\judge{\Gamma}{e}{\tau}{\mualpha}{\alpha}{\mubeta}{\beta}$
in \lambdaF,
then $\judgec{\cpsty{\Gamma}}
             {\cps{e}}
             {\cpsansty{\cpsty{\tau}}
                       {\cpsty{\mualpha}}{\cpsty{\alpha}}
                       {\cpsty{\mubeta}}{\cpsty{\beta}}}$
in \lambdaC.
\end{thm}

\begin{proof}
The proof is by induction on the typing derivation\footnote{The proof is
implemented as function \texttt{cpse} in \texttt{cps.agda}.}.
With the carefully designed rule for case analysis, we can prove the
statement in a straightforward manner, as our type system is directly
derived from the CPS translation.
\end{proof}

From the above theorem, we can further derive the following corollary.

\begin{cor}
An expression $e$ is well-typed in \lambdaF if and only if its CPS 
translation $\cps{e}$ is well-typed in \lambdaC.
\end{cor}

\begin{proof}
The ``if'' direction is trivial: as the type system is directly derived 
from the CPS translation, the statement holds by construction.
The ``only if'' direction is exactly the statement of 
Theorem~\ref{thm:preservationcps}.
\end{proof}

\subsection{Termination of Evaluation}
\label{sec:prop:termination}

Lastly, we conjecture that our type system ensures termination of pure 
programs.

\begin{conj}[Termination]
\label{conj:termination}
If $\judge{\mtenv}{e}{\tau}{\mttrailty}{\alpha}{\mttrailty}{\alpha}$,
then there exists some value $v$ such that $e \reduces v$,
where $\reduces$ is the reflexive, transitive closure of $\reduce$ defined in
Figure~\ref{fig:lambdaf}.
\end{conj}

\noindent The statement is partially witnessed by a CPS interpreter of 
\lambdaF implemented in Agda\footnote{The interpreter is implemented as 
function \texttt{go} in \texttt{lambdaf.agda}.}.
The interpreter shows that any well-typed \lambdaF program evaluates to an 
Agda value under the big-step semantics derived from the CPS translation.
To complete the proof, we must show that the big-step semantics underlying 
the interpreter and the small-step semantics defined for \lambdaF are 
equivalent (\ie, they give us the same value).
Our intuition is that the equivalence holds, but proving this requires 
non-trivial reasoning on continuations, and developing such a proof 
technique is beyond the main scope of this paper.

The termination property, although not fully proved, seems to rely on the 
precise typing of invocation contexts.
Indeed, in the existing type system by Kameyama and Yonezawa~\cite{kameyama-dynamic},
it is possible to write a well-typed program that does not evaluate to a
value.
Here is a non-terminating program adapted from Kameyama and Yonezawa's
example (we write $\seq{e_1}{e_2}$ as an abbreviation of
$\app{(\abs{\_}{e_2})}{e_1}$):

\vspace{-2mm}

\begin{align*}
& \prompt{\seq{(\control{k_1}{\seq{\app{k_1}{1}}{\app{k_1}{1}}})}
              {(\control{k_2}{\seq{\app{k_2}{1}}{\app{k_2}{1}}})}} \\
&= \prompt{\subst{\seq{\app{k_1}{1}}{\app{k_1}{1}}}
                 {k_1}
                 {\abs{x}{\seq{x}{(\control{k_2}{\seq{\app{k_2}{1}}{\app{k_2}{1}}})}}}} \\
&= \prompt{\seq{(\control{k_2}{\seq{\app{k_2}{1}}{\app{k_2}{1}}})}
               {(\app{(\abs{x}{\seq{x}{(\control{k_2}{\seq{\app{k_2}{1}}{\app{k_2}{1}}})}})}{1})}} \\
&= \prompt{\subst{\seq{\app{k_2}{1}}{\app{k_2}{1}}}
                 {k_2}
                 {\abs{y}{\seq{y}{\app{(\abs{x}{\seq{x}{(\control{k_2}{\seq{\app{k_2}{1}}{\app{k_2}{1}}})}})}{1}}}}} \\
&= \prompt{\seq{(\control{k_2}{\seq{\app{k_2}{1}}{\app{k_2}{1}}})}
               {\app{(\abs{y}{\seq{y}{\app{(\abs{x}{\seq{x}{(\control{k_2}{\seq{\app{k_2}{1}}{\app{k_2}{1}}})}})}{1}}})}{1}}} \\
&= ...
\end{align*}

\vspace{2mm}

\noindent We see that the two succeeding invocations of captured
continuations result in duplication of \controltt, leading to a looping
behavior.

The well-typedness of the above program in Kameyama and Yonezawa's type
system is due to the limited expressiveness of trail types.
More precisely, their trail types are mere expression types, which carry
no information about the type of contexts to be composed in the future.
In our type system, on the other hand, trail types explicitly mention the
type of future contexts.
This prevents us from duplicating \controltt infinitely, and allows us to
statically reject the above looping program\footnote{An incomplete
implementation \texttt{exp5} of the non-terminating example is included
in \texttt{lambdaf.agda}}.

%
%
%
%
%
%
%
%
%

\renewcommand{\cps}[1]{\llbracket #1 \rrbracket_C}
\renewcommand{\judgep}[3]{#1 \vdash_p #2 : #3}

\newcommand{\FGCPS}{
\begin{figure}

\leftbf{Syntax of Expression Types}
\begin{align*}
\tau, \alpha, \beta &::=
  ... \mid \arrowtypure{\tau}{\tau}
\end{align*}

\vspace{5mm}

\leftbf{Typing Rules and Selective CPS Translation}

\[
\typing{}
       {\judgeptrans{\Gamma}{n}{\intty}{n}}
       {Int}
\lsep
\typing{x : \tau \in \Gamma}
       {\judgeptrans{\Gamma}{x}{\tau}{x}}
       {Var}
\]

\vspace{4mm}

\[
\typing{\judgeptrans{\extenv{\Gamma}{x}{\tau_1}}{e}{\tau_2}{\epr}}
       {\judgeptrans{\Gamma}{\abs{x}{e}}{\arrowtypure{\tau_1}{\tau_2}}
                    {\abs{x}{\epr}}}
       {PAbs}
\]

\vspace{4mm}

\[
\typing{\judgetrans{\extenv{\Gamma}{x}{\tau_1}}
                  {e}{\tau_2}{\mualpha}{\alpha}{\mubeta}{\beta}{\epr}}
       {\judgeptrans{\Gamma}{\abs{x}{e}}
                    {(\arrowty{\tau_1}{\tau_2}{\mualpha}{\alpha}{\mubeta}{\beta})}
                    {\absthree{x}{\kpr}{\tpr}{\appthree{\epr}{\kpr}{\tpr}}}}
       {IAbs}
\]

\vspace{4mm}

\[
\typing{\judgeptrans{\Gamma}{e_1}{\arrowtypure{\tau_1}{\tau_2}}{\eonepr} \quad
        \judgeptrans{\Gamma}{e_2}{\tau_1}{\etwopr}}
       {\judgeptrans{\Gamma}{\app{e_1}{e_2}}{\tau_2}{\app{\eonepr}{\etwopr}}}
       {PApp}
\]

\vspace{4mm}

\[
\typing{\judgetrans{\Gamma}{e_1}{(\arrowtypure{\tau_1}{\tau_2})}
                   {\mualpha}{\alpha}{\mubeta}{\beta}{\eonepr} \quad
        \judgetrans{\Gamma}{e_2}{\tau_1}{\mugamma}{\gamma}{\mualpha}{\alpha}{\etwopr}}
       {\judgetrans{\Gamma}{\app{e_1}{e_2}}{\tau_2}
                   {\mugamma}{\gamma}{\mubeta}{\beta}
                   {\abs{k}
                        {\app{\eonepr}}
                            {(\abs{v_1}
                                  {\app{\etwopr}
                                       {(\abs{v_2}{\app{k}{(\app{v_1}{v_2})}})}})}}}
       {PIApp}
\]

\vspace{4mm}

\[
\typing{\begin{trgather}
        \judgetrans{\Gamma}{e_1}
                   {(\arrowty{\tau_1}{\tau_2}{\mualpha}{\alpha}{\mubeta}{\beta})}
                   {\mugamma}{\gamma}{\mudelta}{\delta}{\eonepr} \\
        \judgetrans{\Gamma}{e_2}{\tau_1}{\mubeta}{\beta}{\mugamma}{\gamma}{\etwopr}
        \end{trgather}}
       {\judgetrans{\Gamma}{\app{e_1}{e_2}}{\tau_2}
                   {\mualpha}{\alpha}{\mudelta}{\delta}
                   {\abs{k}
                        {\app{\eonepr}}
                            {(\abs{v_1}
                                  {\app{\etwopr}
                                       {(\abs{v_2}{\apptwo{v_1}{v_2}{k}})}})}}}
       {IApp}
\]

\caption{Fine-grained Type System and Selective CPS Translation}
\label{fig:fgcps}
\end{figure}
}

\newcommand{\FGCPSTwo}{
\begin{figure}

\[
\typing{\begin{trgather}
        \judgetrans{\extenv{\Gamma}{c}{\arrowtypure{\tau}{\alpha}}}
                   {e}{\gamma}{\mu_i}{\gammapr}{\mttrailty}{\beta}{\epr} \\
        \idconttype{\gamma}{\mu_i}{\gammapr}
        \end{trgather}}
       {\judgetrans{\Gamma}{\control{c}{e}}{\tau}
                   {\mualpha}{\alpha}{\mualpha}{\beta}
                   {\abstwo{k}{t}
                          {\apptwo{\subst{\epr}
                                         {c}
                                         {\abs{x}{\apptwo{k}{x}{t}}}}
                                  {\idcont}
                                  {\mttrail}}}}
       {PControl}
\]

\vspace{4mm}

\[
\typing{\begin{trgather}
        \judgetrans{\extenv{\Gamma}{c}
                           {\arrowty{\tau}{\tau_1}{\mu_1}{\tauonepr}{\mu_2}{\alpha}}}
                   {e}{\gamma}{\mu_i}{\gammapr}{\mttrailty}{\beta}{\epr} \\
        \idconttype{\gamma}{\mu_i}{\gammapr} \\
        \compatible{(\exttrailty{\tau_1}{\mu_1}{\tauonepr})}{\mu_2}{\mu_0} \\
        \compatible{\mubeta}{\mu_0}{\mualpha}
        \end{trgather}}
       {\begin{trgather}
              \judge{\Gamma}{\control{c}{e}}{\tau}
                   {\mualpha}{\alpha}{\mubeta}{\beta} \\
                   \shade{\trans 
                     \abstwo{k}{t}
                            {\apptwo{\subst{\epr}
                                           {c}
                                           {\absthree{x}{\kpr}{\tpr}
                                                     {\apptwo{k}{x}
                                                             {(\append{t}{(\cons{\kpr}{\tpr})})}}}}
                                    {\idcont}
                                    {\mttrail}}}
       \end{trgather}}
       {IControl}
\]

\vspace{4mm}

\[
\typing{\begin{trgather}
        \judgetrans{\Gamma}{e}{\beta}{\mu_i}{\betapr}{\mttrailty}{\tau}{\epr} \\
        \idconttype{\beta}{\mu_i}{\betapr}
        \end{trgather}}
       {\judgeptrans{\Gamma}{\prompt{e}}{\tau}
                    {\apptwo{\epr}{\idcont}{\mttrail}}}
       {Prompt}
\qquad
\typing{\judgeptrans{\Gamma}{e}{\tau}{\epr}}
       {\judgetrans{\Gamma}{e}{\tau}{\mualpha}{\alpha}{\mualpha}{\alpha}
                   {\abstwo{k}{t}{\apptwo{k}{\epr}{t}}}}
       {Exp}
\]

\caption{Fine-grained Type System and Selective CPS Translation (continued)}
\label{fig:fgcpstwo}
\end{figure}
}

\section{Selective CPS Translation and Fine-grained Type System}
\label{sec:selective}

In the CPS translation defined in Section~\ref{sec:cps}, we converted
every expression into a function that receives a continuation and a trail.
When used as an implementation technique, such a whole-program translation
would blow up the size of programs and lead to poor performance.
A popular solution to this problem is to make the CPS translation
\emph{selective}~\cite{nielsen-selective, kim-assessing, rompf-selective,
asai-selective}, that is, we only convert impure expressions into CPS, and
keep pure expressions in direct style.

In this section, we present a selective CPS translation for \controltt and
\prompttt, as well as a corresponding type system.
Unlike in the previous sections, here we first define the type system
(Section~\ref{sec:selective:type}) \emph{and then} derive the CPS
translation (Section~\ref{sec:selective:cps}).
This is because defining a selective CPS translation requires a purity
analysis, which in turn requires a type system.
We show that, with the distinction between pure and impure expressions,
we can still prove desired properties (Section~\ref{sec:selective:prop}).
Furthermore, if we assume termination of well-typed programs, we can 
macro-express Danvy and Filinski's~\cite{danvy-abstracting} \shifttt 
operator using \controltt and \prompttt (Section~\ref{sec:selective:shift}).

\subsection{Fine-grained Type System}
\label{sec:selective:type}

\FGCPS

\FGCPSTwo

We start by building a fine-grained type system (Figures~\ref{fig:fgcps}
and \ref{fig:fgcpstwo}, ignoring the shaded parts) that is suitable for
defining a selective CPS translation.
As we wish to use the type system to decide whether or not to CPS-translate
an expression, we make a rigorous distinction between pure and impure
expressions.
We do this by introducing a special judgment $\judgep{\Gamma}{e}{\tau}$
for pure expressions, which does not carry answer types or trail types.
This design is borrowed from earlier work on typed control
operators~\cite{asai-polymorphic, kameyama-dynamic}, and reflects the fact
that a pure expression can be evaluated independent of its surrounding
context.
As a corresponding change, we extend the syntax of expression types with a
new arrow type $\arrowtypure{\tau}{\tau}$, which we assign to functions
having a pure body.

Using the pure judgment and function types, we define the typing rules.
There are several possible strategies for designing a fine-grained type
system:

\begin{itemize}
\item Define one rule for each combination of the purity of
  subexpressions~\cite{cong-handling}.
  This allows us to design the most optimized CPS translation, but
  results in a large number of typing rules (for instance, we would
  need $2^3 = 8$ rules for application).

\item Define one rule for each syntactic construct, and in each rule,
  perform a sophisticated computation over the effects of
  subexpressions~\cite{rompf-selective, ishio-verifying}.
  This allows us to keep the type system concise, but makes it hard to
  see how each construct manipulates the answer type and trail type.

\item Define one or more rules for each syntactic construct depending
  on the granularity one wish to achieve, and include a rule for
  converting a pure expression into an impure one~\cite{materzok-subtyping}.
  This allows us to have a reasonably concise type system, at the cost
  of losing some optimization opportunities in the CPS translation.
\end{itemize}

\noindent Here we take the third approach to balance conciseness and
efficiency.

Among different constructs, values and \prompttt are syntactically
classified as pure.
Notice that we have two rules for abstractions, differing in the purity
of the body expression.
In particular, the \rulename{PAbs} rule is necessary for judging certain
applications as pure ones.
As we will see shortly, having pure applications allows us to eliminate
multiple administrative redexes in the CPS translation.
For \prompttt, on the other hand, we only have one rule where the body is
an impure expression.
This does not affect typability, as we can always cast a pure expression
into an impure one via \rulename{Exp}.
The casting introduces some extra redexes in the CPS translation, but
here we prioritize conciseness over efficiency.

An application is pure when the function $e_1$, the argument $e_2$, and
the body of the function are all pure \rulename{PApp}.
The three rules for application, together with \rulename{Exp}, give us
the same expressiveness as having $8$ rules as in the first approach
mentioned above.
Note that, without \rulename{PIApp}, we would not be able to directly
call a function that has a pure body but is itself an impure expression.

A \controltt construct is always judged impure, but it has two rules
differing in the purity of the captured continuation.
The impure rule \rulename{IControl} is identical to the \rulename{Control}
rule from Section~\ref{sec:type}.
The pure rule \rulename{PControl} is simpler than the original rule in that
it imposes no compatibility requirements.
To see why we do not need those requirements, recall that the non-selective
CPS translation of $\control{c}{e}$ extends the given trail.
This is necessary for allowing the invocation context of $c$ to be accessed
by the \controltt operators that reside in $c$.
However, if $c$ is pure, we know that its invocation does not involve any
execution of \controltt.
This means, for a \controltt operator that captures a pure continuation,
we do \emph{not} need to extend the trail.
The absence of the compatibility requirements communicates the fact that
no trail extension happens in this case.
The equal initial and final trail types in the conclusion also reflect
the stability of the trail.
Having \rulename{PControl} could potentially be beneficial for optimizations,
because extending a trail is done by recursively calling the cons function
in the CPS translation.
It would also help us simplify type error messages, as we can assign a pure
function type (that does not mention answer types or trail types) to the
captured continuation.

\subsection{Selective CPS Translation}
\label{sec:selective:cps}

Guided by the fine-grained type system, we define a selective CPS translation
from \lambdaF to \lambdaC (shaded parts of Figures~\ref{fig:fgcps} and
\ref{fig:fgcpstwo}).
The translation keeps pure expressions in direct-style, and turns impure
expressions into a continuation- and trail-taking function.
Unlike the translation defined in Section~\ref{sec:cps}, the selective
translation is defined on the typing derivation, because the information
about purity is not generally available at the level of syntax.
Therefore, we have one translation rule for each typing rule.

Let us take a closer look at the interesting cases.
First, observe the three translation rules for application.
In \rulename{PApp}, we obtain a direct-style application
$\app{\eonepr}{\etwopr}$.
In \rulename{PIApp} and \rulename{IApp}, we obtain a CPS function that uses
the continuation $k$ in a way that is consistent with the purity of the
function body.

Second, compare the translation rules for \controltt.
In \rulename{PControl}, we do \emph{not} extend the trail as we do in
\rulename{IControl}, since we know the captured continuation is pure and
thus cannot manipulate its invocation context.
From an implementation perspective, this means the selective translation
not only produces fewer applications involving continuations and trails,
but also introduces fewer extensions of trails via consing and appending.

Thirdly and lastly, if a pure expression $e$ is judged as impure via
\rulename{Exp}, it is translated to a CPS function that immediately calls
the continuation $k$ with the direct-style translation of $e$ and the given
trail $t$.

\renewcommand{\judgetrans}[8]{\judge{#1}{#2}{#3}{#4}{#5}{#6}{#7} \trans #8}
\renewcommand{\judgeptrans}[4]{\judgep{#1}{#2}{#3} \transp #4}

\subsection{Properties}
\label{sec:selective:prop}

The fine-grained formulation of \lambdaF enjoys similar properties to the
original \lambdaF.
Specifically, the type system is sound, and the selective CPS translation 
is type-preserving.
The termination property should also hold, although we only have a partial 
proof.

\begin{thm}[Type Soundness]
\label{thm:soundfg}
If $\judgep{\mtenv}{e}{\tau}$,
then evaluation of $e$ does not get stuck.
\end{thm}

\begin{proof}
As in the original \lambdaF, we prove type soundness via preservation and
progress\footnote{The preservation theorem is implemented as the
\texttt{Reduce} relation in \texttt{lambdaf-fg-red.agda}.}.
\end{proof}

\begin{thm}[Type Preservation of Selective CPS Translation]
\label{thm:preservationselective}
\begin{enumerate}
\item[]

\item If $\judgeptrans{\Gamma}{e}{\tau}{\epr}$ in \lambdaF,
then $\judgec{\cpsty{\Gamma}}{\epr}{\cpsty{\tau}}$ in \lambdaC.

\item If $\judgetrans{\Gamma}{e}{\tau}{\mualpha}{\alpha}{\mubeta}{\beta}{\epr}$
in \lambdaF,
then $\judgec{\cpsty{\Gamma}}
             {\epr}
             {\cpsansty{\cpsty{\tau}}
                       {\cpsty{\mualpha}}{\cpsty{\alpha}}
                       {\cpsty{\mubeta}}{\cpsty{\beta}}}$
in \lambdaC.
\end{enumerate}
\end{thm}

\begin{proof}
As before, we prove type preservation by induction on the typing
derivation\footnote{The proof is implemented as functions \texttt{cpsp}
and \texttt{cpsi} in \texttt{cps-selective.agda}.}.
Note that we need to extend the type translation with a clause for pure
function types:

\vspace{-2mm}

\begin{align*}
\cpsty{(\arrowtypure{\tau_1}{\tau_2})} &= \arrowtypure{\cpsty{\tau_1}}{\cpsty{\tau_2}}
\end{align*}

\vspace{2mm}

\noindent Notice also that the translation on pure expressions does not
introduce new arrow types, meaning that pure expressions are kept in direct
style.
\end{proof}

\begin{conj}[Termination]
\label{thm:terminationfg}
If $\judgep{\Gamma}{e}{\tau}$,
then there exists some value $v$ such that $e \reduces v$,
where $\reduces$ is the reflexive, transitive closure of $\reduce$ defined in
Figure~\ref{fig:lambdaf}.
\end{conj}

\noindent The statement is again partially witnessed by a CPS interpreter of 
the fine-grained \lambdaF implemented in Agda\footnote{The interpreter is 
implemented as function \texttt{go} in \texttt{lambdaf-fg.agda}.}.

\subsection{Encoding of \shifttt}
\label{sec:selective:shift}

Kameyama and Yonezawa~\cite{kameyama-dynamic} show that their type system
allows a typed encoding of \shifttt in terms of \controltt and \prompttt.
We show that this is possible in our fine-grained calculus as well, 
assuming that the termination conjecture actually holds.

Let us begin by comparing the reduction rules of \shifttt and \controltt:

\vspace{-2mm}

\begin{align*}
E[\prompt{F[\shift{k}{e}]}] &\reduce E[\reset{\subst{e}{k}{\abs{x}{\reset{F[x]}}}}] \\
E[\prompt{F[\control{k}{e}]}] &\reduce E[\prompt{\subst{e}{k}{\abs{x}{F[x]}}}]
\end{align*}

\vspace{2mm}

\noindent We see that the two rules differ in the presence of the delimiter
surrounding the captured continuation.
In other words, a continuation captured by \shifttt is always pure, while
a continuation captured by \controltt may be impure.

There is a straightforward encoding of \shifttt by \controltt and
\prompttt~\cite{biernacki-simple}: we simply wrap the continuation captured
by \controltt around \prompttt.

\vspace{-2mm}

\begin{align*}
\shift{k}{e} &= \control{\kpr}{\subst{e}{k}{\abs{x}{\prompt{\app{\kpr}{x}}}}}
\end{align*}

\vspace{2mm}

The encoding is well-typed in the fine-grained \lambdaF\footnote{The
encoding is implemented as \texttt{shift} in \texttt{lambdaf-fg.agda}.}.
Specifically, the \controltt construct can be typed by \rulename{PControl},
because the captured continuation is a pure function.
The absence of compatibility requirements, as well as the equal initial and
final trail types, are consistent with the semantics of the \shifttt
operator: it does not modify the trail as general \controltt may do.

Note that the encoding is \emph{not} well-typed in the original \lambdaF.
The reason is that rule \rulename{Control} requires a compatibility
relation that can never hold.
Let us look at the derivation of the \controltt construct:

\im{
\typing{\begin{trgather}
        \judge{\extenv{\Gamma}{\kpr}
                      {\arrowty{\tau}{\alpha}{\mttrailty}{\alpha}{\mttrailty}{\alpha}}}
              {\subst{e}{k}{\abs{x}{\prompt{\app{\kpr}{x}}}}}
              {\gamma}{\mu_i}{\gammapr}{\mttrailty}{\beta} \\
        \idconttype{\gamma}{\mu_i}{\gammapr} \\
        \compatible{(\exttrailty{\alpha}{\mttrailty}{\alpha})}{\mttrailty}{\mu_0} \\
        \compatible{\mualpha}{\mu_0}{\mualpha}
        \end{trgather}}
       {\judge{\Gamma}{\control{\kpr}{\subst{e}{k}{\abs{x}{\prompt{\app{\kpr}{x}}}}}}{\tau}{\mualpha}{\alpha}{\mualpha}{\beta}}
       {Control}
}

\noindent The instantiation of answer types and trail types is drawn from
the typing constraints of \controltt and \prompttt.
By the definition of \compatiblesf, we know the trail type $\mu_0$ in the
first compatibility relation must be $\exttrailty{\alpha}{\mttrailty}{\alpha}$.
By substituting this trail type for the $\mu_0$ in the second compatibility
relation, we obtain
$\compatible{\mualpha}{(\exttrailty{\alpha}{\mttrailty}{\alpha})}{\mualpha}$.
This relation does not hold, as we prove in our Agda formalization\footnote{
The impossibility proof \texttt{c} is included in \texttt{lambdaf.agda}.}.
As a consequence, we cannot close the above derivation.

In the fine-grained \lambdaF, we can in fact simplify the encoding as
follows:

\vspace{-2mm}

\begin{align*}
\shift{k}{e} &= \control{k}{e} \quad \text{where $k$ is pure}
\end{align*}

\vspace{2mm}

\noindent The simplified encoding says: a \controltt operator that captures
a pure continuation behaves the same as a \shifttt operator.
This can be proved by using the fact that the \prompttt operator is no-op
when its body is a pure expression.
Specifically, if $e$ is pure, the expression $\prompt{e}$ eventually
evaluates to $\prompt{v}$ (under the assumption that termination holds), and 
then to $v$ by the reduction rule \rulename{$\promptsym$}.
Using this fact, we can rewrite the original encoding to
$\control{\kpr}{\subst{e}{k}{\abs{x}{\app{\kpr}{x}}}}$,
and then to $\shift{k}{e}$ by $\eta$-reduction and renaming.

\section{Formalization in Agda}
\label{sec:agda}

We have formalized \lambdaF, \lambdaC, and the CPS translations in the
Agda proof assistant.
In this section, we give the reader a high-level idea of how we formalized
these artifacts.
For the full formalization, we invite the reader to visit the following
GitHub repository:

\vspace{2mm}

\begin{center}
\url{https://github.com/YouyouCong/lmcs-artifact}
\end{center}

\paragraph{Defining Syntax and Typing Rules}

Let us begin with the formalization of (original) \lambdaF.
Following the recipe of Altenkirch and Reus~\cite{altenkirch-monadic}, we
define \lambdaF as an \emph{intrinsically-typed} language.
That is, we define the syntax and typing at once, and thus guarantee
that we can only construct well-typed expressions.

Before defining expressions, we define expression types and trail types
as datatypes \texttt{\Ty} and \texttt{\Tr}.

\vspace{4mm}

\begin{alltt}
\data \Ty : \Set
\data \Tr : \Set

\data \Ty \where
  \Nat         : \Ty
  \Bool        : \Ty
  \Str         : \Ty
  \Arr{\gdummy}{\gdummy}{\gdummy}{\gdummy}{\gdummy}{\gdummy} : \Ty \arr \Ty \arr \Tr \arr \Ty \arr \Tr \arr \Ty \arr \Ty

\data \Tr \where
  \Bul      : \Tr
  \Ext{\gdummy}{\gdummy}{\gdummy} : \Ty \arr \Tr \arr \Ty \arr \Tr
\end{alltt}

\vspace{4mm}

\noindent The two datatypes \texttt{\Ty} and \texttt{\Tr} are declared
as an inhabitant of \texttt{\Set}, which represents the type of types.
The separation between datatype and constructor signatures indicates that
\texttt{\Ty} and \texttt{\Tr} are defined by mutual induction.
In \texttt{\Ty}, we include three base types that are necessary for 
expressing the two examples from Section~\ref{sec:control}.

Using \texttt{\Ty} and \texttt{\Tr}, we define values and expressions.

\vspace{4mm}

\begin{alltt}
\data \Val{\bdummy}{\bdummy}  (var : \Ty \arr \Set) : \Ty \arr \Set
\data \Exp{\bdummy}{\bdummy}{\bdummy}{\bdummy}{\bdummy}{\bdummy}  (var : \Ty \arr \Set) : \Ty \arr \Tr \arr \Ty \arr \Tr \arr \Ty \arr \Set

\data \Val{\bdummy}{\bdummy}  \where
  \Var : \{\tauty : \Ty\} \arr var \tauty \arr \Val{var}{\tauty}
  ...

\data \Exp{\bdummy}{\bdummy}{\bdummy}{\bdummy}{\bdummy}{\bdummy}  \where
  \Valcon    : \{\tauty \alphaty : \Ty\} \{\mualphaty : \Tr\} \arr
           \Val{var}{\tauty} \arr \Exp{var}{\tauty}{\mualphaty}{\alphaty}{\mualphaty}{\alphaty}
  \App    : \{\tauonety \tautwoty \alphaty \betaty \gammaty \deltaty : \Ty\} \{\mualphaty \mubetaty \mugammaty \mudeltaty : \Tr\} \arr
           \Exp{var}{(\Arr{\tauonety}{\tautwoty}{\mualphaty}{\alphaty}{\mubetaty}{\betaty})}{\mugammaty}{\gammaty}{\mudeltaty}{\deltaty} \arr
           \Exp{var}{\tauonety}{\mubetaty}{\betaty}{\mugammaty}{\gammaty} \arr
           \Exp{var}{\tautwoty}{\mualphaty}{\alphaty}{\mudeltaty}{\deltaty}
  ...
  \Prompt : \{\tauty \alphaty \betaty \betaprty : \Ty\} \{\muity \mualphaty : \Tr\} \arr
           \idconttypefunc \betaty \muity \betaprty \arr
           \Exp{var}{\betaty}{\muity}{\betaprty}{\Bul}{\tauty} \arr
           \Exp{var}{\tauty}{\mualphaty}{\alphaty}{\mualphaty}{\alphaty}
\end{alltt}

\vspace{4mm}

\noindent The type of expressions can be read as an encoding of the
typing judgment used in the typing rules.
The only gap between the two lies in the treatment of variables.
In our formalization, we adopt \emph{parametric higher-order abstract
syntax (PHOAS)}~\cite{chlipala-phoas}.
The technique allows us to use meta language binders to manage object
language bindings, without breaking the positivity of the datatype.
In our case, we parameterize \texttt{\func{Val}} and \texttt{\func{Exp}}
over \texttt{var}, which is a function from \lambdaF types to Agda
types\footnote{The \texttt{var} argument is shared among all constructors
of \texttt{\func{Val}} and \texttt{\func{Exp}}.
For this reason, we declare \texttt{var} as a \emph{parameter} (by
writing it before the colon), and thus avoid introducing it for every
constructor signature.
In contrast, the type and trail type arguments vary across constructors.
Therefore, we declare them as \emph{indices} (by writing them after the
colon), and introduce them for every constructor signature as implicit
arguments (by enclosing them in curly brackets).}.

The constructors of \texttt{\func{Val}} and \texttt{\func{Exp}} represent
the typing rules of \lambdaF.
For example, the \texttt{\App} constructor takes in two arguments
corresponding to the two premises in the typing rule \rulename{App}.
As a different example, the \texttt{\Prompt} constructor demands a proof
of the \texttt{\idconttypefunc} relation, which is defined as an Agda
function returning a type.

\paragraph{Defining Reduction Rules}

Having defined the syntax and typing rules, we define the reduction rules.

\vspace{4mm}

\begin{alltt}
\data \Reduce \{var : \Ty \arr \Set\} :
            \{\tauty \alphaty \betaty : \Ty\} \{\mualphaty \mubetaty : \Tr\} \arr
            \Exp{var}{\tauty}{\mualphaty}{\alphaty}{\mubetaty}{\betaty} \arr
            \Exp{var}{\tauty}{\mualphaty}{\alphaty}{\mubetaty}{\betaty} \arr \Set \where
  ...
  \RPrompt : \{\tauty : \Ty\} \{v : \Val{var}{\tauty}\} \arr
            \Reduce (\Prompt (\Valcon v)) (\Valcon v)
\end{alltt}

\vspace{4mm}

\noindent The type \texttt{\Reduce e e'} means $e \reduce \epr$, and has
one constructor for each reduction rule in Figure~\ref{fig:lambdaf}.
For instance, \texttt{\RPrompt} represents the rule $\prompt{v} \reduce v$.
Notice that the two arguments to \texttt{\Reduce} have the same type.
This means, if we can define all the reduction rules as a constructor of
\texttt{\Reduce}, then we can be sure that reducing a well-typed expression
produces an expression of the same type.
Therefore, the definition of \texttt{\Reduce} is essentially the proof of
the preservation theorem (Theorem~\ref{thm:preservation}).

\paragraph{Defining CPS Translation}

Now we proceed to the CPS translation.
We formalize the target calculus \lambdaC in a similar way to \lambdaF,
and define the CPS translation of types and expressions.

\vspace{4mm}

\begin{alltt}
\cpstau{\bdummy} : \Ty \arr \CTy
\cpsmu{\bdummy} : \Tr \arr \CTy

\cpsv : \{var : \CTy \arr \Set\} \{\tauty : \Ty\} \arr
       \Val{var \bcirc \cpstau{\bdummy}}{\tauty} \arr
       \CVal{var}{\cpstau{\tauty}}
\cpse : \{var : \CTy \arr \Set\} \{\tauty \alphaty \betaty : \Ty\} \{\mualphaty \mubetaty : \Tr\} \arr
       \Exp{var \bcirc \cpstau{\bdummy}}{\tauty}{\mualphaty}{\alphaty}{\mubetaty}{\betaty} \arr
       \CVal{var}{((\cpstau{\tauty} \darr \cpsmu{\mualphaty} \darr \cpstau{\alphaty}) \darr \cpsmu{\betaty} \darr \cpstau{\betaty})}
\end{alltt}

\vspace{4mm}

\noindent In the above definition, \texttt{\CTy} is the type of \lambdaC
types, and \texttt{\func{CVal}} is the type of \lambdaC values.
As we can see, the signature of \texttt{\cpse} is exactly the statement
of the type preservation theorem (Theorem~\ref{thm:preservationcps}).
Thus, the definition of \texttt{\cpse} serves as a proof of type
preservation.

\section{Related Work}
\label{sec:related}

\paragraph{Variations of Control Operators}
There are four variants of delimited control operators in the style of
\controltt, differing in whether the operator keeps the surrounding delimiter,
and whether it inserts a delimiter into the captured
continuation~\cite{dyvbig-monadic}.
We have seen that the absence of the latter delimiter makes it possible to
capture invocation contexts.
The lack of the former delimiter, on the other hand, gives us access to
\emph{metacontexts}, i.e., contexts outside of the innermost delimiter.
In the CPS translation, metacontexts are represented using a stack-like data
structure, which is also called a trail in the
literature~\cite{materzok-subtyping}.

As an orthogonal variation, some implementations of control operators
provide \emph{prompt tags}~\cite{gunter-cupto, kiselyov-delimcc}, allowing
the programmer to specify the association between the control operator and
the delimiter.
Since tagged control operators make it possible to skip intervening
delimiters, their CPS translation also involves a form of trails, containing
the delimited contexts within individual delimiters.

\paragraph{Type Systems for Control Operators}
The CPS-based approach to designing type systems has been applied to
three variants of delimited control operators, namely
\shifttt/\resettt \cite{danvy-context, asai-polymorphic},
\controltt/\prompttt~\cite{kameyama-dynamic}, and
\shiftztt/\resetztt~\cite{materzok-subtyping}.
Among them, \shiftztt can capture metacontexts, hence its type system
maintains a list of answer types, representing the return types of nesting
delimiters~\cite{materzok-subtyping}.

The CPS approach gives us by default a type system where every expression
is effectful, as in the original \lambdaF.
Such a type system is however impractical, as it rejects many programs
that are in fact type-safe.
For instance, we showed that the encoding of \shifttt by
\controltt/\prompttt is ill-typed if \controltt can only ever capture an
impure continuation.
It is also known that implementing the list prefix function using
\shifttt/\resettt requires continuations to be \emph{answer-type
polymorphic}~\cite{asai-polymorphic}, which basically means ``pure''.
For this reason, many type systems for control operators employ a purity
distinction as in the fine-grained \lambdaF~\cite{asai-polymorphic,
kameyama-dynamic, materzok-subtyping}.


\paragraph{Selective CPS Translations and Fine-grained Type Systems}
The idea of selectively CPS-translating programs based on a fine-grained
type system has long been studied as a compilation technique for effectful
languages.
For example, Kim et al.~\cite{kim-assessing} use a selective CPS
translation to reduce the overhead of ML exceptions.
As a different example, Asai and Uehara~\cite{asai-selective} implement
the \shifttt and \resettt operators via a selective CPS translation.
A particularly interesting result of the latter work is that, by treating
\shifttt-captured continuations as pure functions, we can significantly
speed-up non-deterministic programs (such as the N-Queen puzzle).
In the case of \controltt, the impact of pure continuations may be even
more remarkable, because they not only reduce continuation parameters,
but also eliminate consing and appending of trails.

\paragraph{Algebraic Effects and Handlers}
In the past decade, algebraic effects and
handlers~\cite{bauer-tutorial, plotkin-handler} have become a popular
tool for handling delimited continuations.
A prominent feature of effect handlers is that a captured continuation is
used at the delimiter site.
This makes it unnecessary to keep track of answer types in the type
system, as we can decide within a handler whether the use of a continuation
is consistent with the actual context.
The irrelevance of answer types in turn makes the connection between the
type system and CPS translation looser.
Indeed, type systems of effect handlers~\cite{bauer-effsys, kammar-handler}
existed before their CPS
semantics~\cite{leijen-cps, hillerstrom-cps, hillerstrom-jfp}.
Also, type-preserving CPS translation of effect handlers is still an open
problem in the community~\cite{hillerstrom-jfp}.

\section{Conclusion and Future Work}
\label{sec:conclusion}

In this paper, we show how to derive a fully expressive type system for 
the \controltt and \prompttt operators.
The main idea is to identify all the typing constraints by analyzing a CPS
translation, where trails are represented as a composition of functions.

The present study is part of a long-term project on formalizing delimited
control facilities whose theory is not yet fully developed.
In the rest of this section, we describe several directions for future work.

\paragraph{Polymorphism and Type Inference/Checking}
Having established a principle for monomorphic typing, a natural next step
is to extend the calculus with polymorphism.
The combination of polymorphism and effects is a delicate issue, and often
requires a value or purity restriction to restore type
soundness~\cite{harper-polymorphic, asai-polymorphic}.
In the presence of \controltt/\prompttt, we must further support abstraction
over trail types, because otherwise we would not be able to call a function
or continuation in different contexts.

After adding polymorphism, we would like to design an algorithm for type
inference and type checking.
We conjecture that answer types can be left implicit in the user program,
because it is the case in a calculus featuring \shifttt and
\resettt~\cite{asai-polymorphic}.
On the other hand, we anticipate that some of the trail types need to be
explicitly given by the user, as it does not seem always possible to synthesize
the intermediate trail types ($\mu_0$, $\exttrailty{\tau_1}{\mu_1}{\tauonepr}$,
and $\mu_2$) in the \rulename{Control} rule.

\paragraph{Implementation}
With polymorphism and a type checking/inference algorithm, we are ready to
develop a practical implementation of our \controltt/\prompttt calculus.
There are several approaches to implementing the dynamic behavior of the
control operators.
One way is to use the CPS translation defined in this paper, which eliminates
the need for special runtime support.
A more direct approach is to implement Fujii and Asai's~\cite{fujii-derivation}
compiler, which is derived from a CPS interpreter by applying various
program transformations.
Once we have a language that we can play with, we will assess the usability
of our language from the viewpoints of expressiveness and efficiency.

\paragraph{Equational Theory and Reflection}
So far, the semantics of \controltt and \prompttt has been given as a CPS
translation and an abstract machine~\cite{shan-simulation, biernacki-toplas}.
In future work, we intend to develop an alternative semantics in the form of
an \emph{equational theory}.
Such a semantics exists for \shifttt/\resettt~\cite{kameyama-axiom} and
\shiftztt/\resetztt~\cite{materzok-axiomatizing}, and is especially useful
for compilation.
For instance, it enables converting an optimization in a CPS compiler into
a rewrite in a direct-style program~\cite{sabry-reasoning}.

An equational theory can further be refined into a
\emph{reflection}~\cite{sabry-reflection}, where direct-style and CPS
expressions are related by reduction rather than equation.
Given the recent advance in the reflection for delimited control
operators~\cite{biernacki-ccs, biernacki-stacked}, we would naturally
ask whether it is possible to extend the results to \controltt/\prompttt.
To answer this question, we need to first define a version of CPS
translation that does not yield administrative redexes~\cite{plotkin-cps,
danvy-onepass}, including the applications arising from consing and
appending of trails.

\paragraph{Control0/Prompt0 and Shallow Effect Handlers}
There is a variation of \controltt and \prompttt, called \controlztt and
\promptztt, that remove the matching delimiter upon capturing of a
continuation.
We are currently formalizing a typed calculus of \controlztt/\promptztt,
by combining the insights from the present work and previous study on
\shiftztt/\resetztt~\cite{materzok-subtyping}.
As shown by Pir\'og et al.~\cite{pirog-typedeq}, there exists a pair of
macro translations \cite{felleisen-macro} between \controlztt/\promptztt
and \emph{shallow effect handlers}~\cite{hillerstrom-shallow}.
This suggests that a good understanding of \controlztt/\promptztt would 
serve as a stepping stone to a better implementation of shallow handlers.
For example, establishing equational axioms for \controlztt/\promptztt 
would help us establish similar axioms for shallow handlers, which would 
in turn be used to optimize languages with support for shallow handlers,
such as Frank~\cite{lindley-frank} and Koka~\cite{leijen-koka}.

\bibliographystyle{alphaurl}
\bibliography{lmcs}

\newcommand{\etalchar}[1]{$^{#1}$}
\begin{thebibliography}{WZD{\etalchar{+}}19}

\bibitem[ABDM03]{ager-functional}
Mads~Sig Ager, Dariusz Biernacki, Olivier Danvy, and Jan Midtgaard.
\newblock A functional correspondence between evaluators and abstract machines.
\newblock In {\em Proceedings of the 5th ACM SIGPLAN International Conference
  on Principles and Practice of Declaritive Programming}, PPDP '03, pages
  8--19, New York, NY, USA, 2003. ACM.
\newblock \href {https://doi.org/10.1145/888251.888254}
  {\path{doi:10.1145/888251.888254}}.

\bibitem[AK07]{asai-polymorphic}
Kenichi Asai and Yukiyoshi Kameyama.
\newblock Polymorphic delimited continuations.
\newblock In {\em Proceedings of the 5th Asian Conference on Programming
  Languages and Systems}, APLAS '07, pages 239--254, Berlin, Heidelberg, 2007.
  Springer-Verlag.

\bibitem[AR99]{altenkirch-monadic}
Thorsten Altenkirch and Bernhard Reus.
\newblock Monadic presentations of lambda terms using generalized inductive
  types.
\newblock In {\em International Workshop on Computer Science Logic}, pages
  453--468. Springer, 1999.

\bibitem[AU17]{asai-selective}
Kenichi Asai and Chihiro Uehara.
\newblock Selective {CPS} transformation for shift and reset.
\newblock In {\em Proceedings of the ACM SIGPLAN Workshop on Partial Evaluation
  and Program Manipulation}, PEPM '18, pages 40--52, New York, NY, USA, dec
  2017. ACM.
\newblock \href {https://doi.org/10.1145/3162069} {\path{doi:10.1145/3162069}}.

\bibitem[BBD05]{biernacka-operational}
Malgorzata Biernacka, Dariusz Biernacki, and Olivier Danvy.
\newblock An operational foundation for delimited continuations in the {CPS}
  hierarchy.
\newblock {\em Logical Methods in Computer Science}, 1, 2005.

\bibitem[BD05]{biernacki-simple}
Dariusz Biernacki and Olivier Danvy.
\newblock A simple proof of a folklore theorem about delimited control.
\newblock {\em BRICS Report Series}, 12(25), 2005.

\bibitem[BDM06]{biernacki-brics}
Dariusz Biernacki, Olivier Danvy, and Kevin Millikin.
\newblock A dynamic continuation-passing style for dynamic delimited
  continuations.
\newblock {\em BRICS Report Series}, 13(15), 2006.

\bibitem[BDM15]{biernacki-toplas}
Dariusz Biernacki, Olivier Danvy, and Kevin Millikin.
\newblock A dynamic continuation-passing style for dynamic delimited
  continuations.
\newblock {\em ACM Trans. Program. Lang. Syst.}, 38(1), October 2015.
\newblock \href {https://doi.org/10.1145/2794078} {\path{doi:10.1145/2794078}}.

\bibitem[BDS06]{biernacki-bft}
Dariusz Biernacki, Olivier Danvy, and Chung-chieh Shan.
\newblock On the static and dynamic extents of delimited continuations.
\newblock {\em Science of Computer Programming}, 60(3):274--297, 2006.

\bibitem[BP13]{bauer-effsys}
Andrej Bauer and Matija Pretnar.
\newblock An effect system for algebraic effects and handlers.
\newblock In Reiko Heckel and Stefan Milius, editors, {\em Algebra and
  Coalgebra in Computer Science}, pages 1--16, Berlin, Heidelberg, 2013.
  Springer Berlin Heidelberg.

\bibitem[BP15]{bauer-tutorial}
Andrej Bauer and Matija Pretnar.
\newblock Programming with algebraic effects and handlers.
\newblock {\em Journal of Logical and Algebraic Methods in Programming},
  84(1):108--123, 2015.

\bibitem[BPS20]{biernacki-ccs}
Dariusz Biernacki, Mateusz Pyzik, and Filip Sieczkowski.
\newblock A reflection on continuation-composing style.
\newblock In {\em Proceedings of the 5th International Conference on Formal
  Structures for Computation and Deduction}, FSCD '20. Schloss
  Dagstuhl--Leibniz-Zentrum fuer Informatik, 2020.

\bibitem[BPS21]{biernacki-stacked}
Dariusz Biernacki, Mateusz Pyzik, and Filip Sieczkowski.
\newblock Reflecting stacked continuations in a fine-grained direct-style
  reduction theory.
\newblock In {\em Proceedings of the 23rd International Symposium on Principles
  and Practice of Declarative Programming}, PPDP '21. ACM, 2021.

\bibitem[CA18]{cong-handling}
Youyou Cong and Kenichi Asai.
\newblock Handling delimited continuations with dependent types.
\newblock {\em Proc. ACM Program. Lang.}, 2(ICFP):69:1--69:31, July 2018.
\newblock \href {https://doi.org/10.1145/3236764} {\path{doi:10.1145/3236764}}.

\bibitem[Chl08]{chlipala-phoas}
Adam Chlipala.
\newblock Parametric higher-order abstract syntax for mechanized semantics.
\newblock In {\em Proceedings of the 13th ACM SIGPLAN International Conference
  on Functional Programming}, ICFP '08, page 143–156, New York, NY, USA,
  2008. Association for Computing Machinery.
\newblock \href {https://doi.org/10.1145/1411204.1411226}
  {\path{doi:10.1145/1411204.1411226}}.

\bibitem[CIHA21]{cong-fscd}
Youyou Cong, Chiaki Ishio, Kaho Honda, and Kenichi Asai.
\newblock {A Functional Abstraction of Typed Invocation Contexts}.
\newblock In Naoki Kobayashi, editor, {\em 6th International Conference on
  Formal Structures for Computation and Deduction (FSCD 2021)}, volume 195 of
  {\em Leibniz International Proceedings in Informatics (LIPIcs)}, pages
  12:1--12:18, Dagstuhl, Germany, 2021. Schloss Dagstuhl -- Leibniz-Zentrum
  f{\"u}r Informatik.
\newblock URL: \url{https://drops.dagstuhl.de/opus/volltexte/2021/14250}, \href
  {https://doi.org/10.4230/LIPIcs.FSCD.2021.12}
  {\path{doi:10.4230/LIPIcs.FSCD.2021.12}}.

\bibitem[Coq92]{coquand-pattern}
Thierry Coquand.
\newblock Pattern matching with dependent types.
\newblock In {\em Proceedings of the Third Workshop on Logical Frameworks},
  1992.

\bibitem[Dan96]{danvy-tdpe}
Olivier Danvy.
\newblock Type-directed partial evaluation.
\newblock In {\em Proceedings of the 23rd ACM SIGPLAN-SIGACT Symposium on
  Principles of Programming Languages}, POPL '96, pages 242--257. ACM, 1996.

\bibitem[DF89]{danvy-context}
Olivier Danvy and Andrzej Filinski.
\newblock A functional abstraction of typed contexts.
\newblock BRICS 89/12, August 1989.

\bibitem[DF90]{danvy-abstracting}
Olivier Danvy and Andrzej Filinski.
\newblock Abstracting control.
\newblock In {\em Proceedings of the 1990 ACM conference on LISP and functional
  programming}, pages 151--160. ACM, 1990.

\bibitem[DN03]{danvy-onepass}
Olivier Danvy and Lasse~R Nielsen.
\newblock A first-order one-pass cps transformation.
\newblock {\em Theoretical Computer Science}, 308(1-3):239--257, 2003.

\bibitem[DPJS07]{dyvbig-monadic}
R.~Kent Dyvbig, Simon Peyton~Jones, and Amr Sabry.
\newblock A monadic framework for delimited continuations.
\newblock {\em J. Funct. Program.}, 17(6):687--730, November 2007.
\newblock \href {https://doi.org/10.1017/S0956796807006259}
  {\path{doi:10.1017/S0956796807006259}}.

\bibitem[FA21]{fujii-derivation}
Maika Fujii and Kenichi Asai.
\newblock {Derivation of a Virtual Machine For Four Variants of
  Delimited-Control Operators}.
\newblock In Naoki Kobayashi, editor, {\em 6th International Conference on
  Formal Structures for Computation and Deduction (FSCD 2021)}, volume 195 of
  {\em Leibniz International Proceedings in Informatics (LIPIcs)}, pages
  16:1--16:19, Dagstuhl, Germany, 2021. Schloss Dagstuhl -- Leibniz-Zentrum
  f{\"u}r Informatik.
\newblock URL: \url{https://drops.dagstuhl.de/opus/volltexte/2021/14254}, \href
  {https://doi.org/10.4230/LIPIcs.FSCD.2021.16}
  {\path{doi:10.4230/LIPIcs.FSCD.2021.16}}.

\bibitem[Fel88]{felleisen-prompt}
Mattias Felleisen.
\newblock The theory and practice of first-class prompts.
\newblock In {\em Proceedings of the 15th ACM SIGPLAN-SIGACT Symposium on
  Principles of Programming Languages}, POPL '88, pages 180--190, New York, NY,
  USA, 1988. ACM.
\newblock \href {https://doi.org/10.1145/73560.73576}
  {\path{doi:10.1145/73560.73576}}.

\bibitem[Fel91]{felleisen-macro}
Matthias Felleisen.
\newblock On the expressive power of programming languages.
\newblock In {\em Selected Papers from the Symposium on 3rd European Symposium
  on Programming}, ESOP '90, pages 35--75, New York, NY, USA, 1991. Elsevier
  North-Holland, Inc.

\bibitem[Fil94]{filinski-representing}
Andrzej Filinski.
\newblock Representing monads.
\newblock In {\em Proceedings of the 21st ACM SIGPLAN-SIGACT Symposium on
  Principles of Programming Languages}, POPL '94, pages 446--457, New York, NY,
  USA, 1994. ACM.
\newblock \href {https://doi.org/10.1145/174675.178047}
  {\path{doi:10.1145/174675.178047}}.

\bibitem[FKLP17]{forster-macro}
Yannick Forster, Ohad Kammar, Sam Lindley, and Matija Pretnar.
\newblock On the expressive power of user-defined effects: Effect handlers,
  monadic reflection, delimited control.
\newblock {\em Proc. ACM Program. Lang.}, 1(ICFP):13:1--13:29, August 2017.
\newblock \href {https://doi.org/10.1145/3110257} {\path{doi:10.1145/3110257}}.

\bibitem[GRR95]{gunter-cupto}
Carl~A. Gunter, Didier R{\'e}my, and Jon~G. Riecke.
\newblock A generalization of exceptions and control in {ML}-like languages.
\newblock In {\em Proceedings of the Seventh International Conference on
  Functional Programming Languages and Computer Architecture}, FPCA '95, pages
  12--23, New York, NY, USA, 1995. ACM.
\newblock \href {https://doi.org/10.1145/224164.224173}
  {\path{doi:10.1145/224164.224173}}.

\bibitem[HL93]{harper-polymorphic}
Robert Harper and Mark Lillibridge.
\newblock Polymorphic type assignment and cps conversion.
\newblock {\em LISP and Symbolic Computation}, 6(3):361--379, 1993.

\bibitem[HL18]{hillerstrom-shallow}
Daniel Hillerstr{\"o}m and Sam Lindley.
\newblock Shallow effect handlers.
\newblock In {\em Asian Symposium on Programming Languages and Systems}, APLAS
  '18, pages 415--435. Springer, 2018.

\bibitem[HLA20]{hillerstrom-jfp}
Daniel Hillerstr\"{o}m, Sam Lindley, and Robert Atkey.
\newblock Effect handlers via generalised continuations.
\newblock {\em Journal of Functional Programming}, 30, 2020.
\newblock \href {https://doi.org/10.1017/S0956796820000040}
  {\path{doi:10.1017/S0956796820000040}}.

\bibitem[HLAS17]{hillerstrom-cps}
Daniel Hillerstr{\"o}m, Sam Lindley, Robert Atkey, and KC~Sivaramakrishnan.
\newblock Continuation passing style for effect handlers.
\newblock In {\em Proceedings of 2nd International Conference on Formal
  Structures for Computation and Deduction}, FSCD '17, pages 18:1--18:19.
  Schloss Dagstuhl--Leibniz-Zentrum fuer Informatik, 2017.

\bibitem[IA19]{ishio-verifying}
Chiaki Ishio and Kenichi Asai.
\newblock Verifying selective cps transformation for shift and reset.
\newblock In William Bowman and Ronald Garcia, editors, {\em International
  Symposium on Trends in Functional Programming}, volume 12053 of {\em Lecture
  Notes in Computer Science (LNCS)}, pages 38--57. Springer, 2019.

\bibitem[KH03]{kameyama-axiom}
Yukiyoshi Kameyama and Masahito Hasegawa.
\newblock A sound and complete axiomatization of delimited continuations.
\newblock In {\em Proceedings of the 8th ACM SIGPLAN International Conference
  on Functional Programming}, ICFP '03, pages 177--188. ACM, 2003.

\bibitem[Kis12]{kiselyov-delimcc}
Oleg Kiselyov.
\newblock Delimited control in ocaml, abstractly and concretely.
\newblock {\em Theor. Comput. Sci.}, 435:56–76, June 2012.
\newblock \href {https://doi.org/10.1016/j.tcs.2012.02.025}
  {\path{doi:10.1016/j.tcs.2012.02.025}}.

\bibitem[KLO13]{kammar-handler}
Ohad Kammar, Sam Lindley, and Nicolas Oury.
\newblock Handlers in action.
\newblock In {\em Proceedings of the 18th ACM SIGPLAN International Conference
  on Functional Programming}, ICFP '13, pages 145--158, New York, NY, USA,
  2013. ACM.
\newblock \href {https://doi.org/10.1145/2500365.2500590}
  {\path{doi:10.1145/2500365.2500590}}.

\bibitem[KS16]{kiselyov-eff}
Oleg Kiselyov and K.~C. Sivaramakrishnan.
\newblock Eff directly in ocaml.
\newblock In {\em ML Workshop}, 2016.

\bibitem[KY08]{kameyama-dynamic}
Yukiyoshi Kameyama and Takuo Yonezawa.
\newblock Typed dynamic control operators for delimited continuations.
\newblock In {\em International Symposium on Functional and Logic Programming},
  FLOPS '08, pages 239--254. Springer, 2008.

\bibitem[KYD98]{kim-assessing}
Jung-taek Kim, Kwangkeun Yi, and Olivier Danvy.
\newblock Assessing the overhead of {ML} exceptions by selective {CPS}.
\newblock In {\em Proceedings of the 1998 ACM SIGPLAN Workshop on ML}, 1998.

\bibitem[Lei14]{leijen-koka}
Daan Leijen.
\newblock Koka: Programming with row polymorphic effect types.
\newblock In {\em 5th Workshop on Mathematically Structured Functional
  Programming}, MSFP '14, 2014.
\newblock \href {https://doi.org/10.4204/EPTCS.153.8}
  {\path{doi:10.4204/EPTCS.153.8}}.

\bibitem[Lei17]{leijen-cps}
Daan Leijen.
\newblock Type directed compilation of row-typed algebraic effects.
\newblock In {\em Proceedings of the 44th ACM SIGPLAN Symposium on Principles
  of Programming Languages}, POPL '17, pages 486--499, New York, NY, USA, 2017.
  ACM.
\newblock \href {https://doi.org/10.1145/3009837.3009872}
  {\path{doi:10.1145/3009837.3009872}}.

\bibitem[LMM17]{lindley-frank}
Sam Lindley, Conor McBride, and Craig McLaughlin.
\newblock Do be do be do.
\newblock In {\em Proceedings of the 44th ACM SIGPLAN Symposium on Principles
  of Programming Languages}, POPL '17, pages 500--514, New York, NY, USA, 2017.
  ACM.
\newblock \href {https://doi.org/10.1145/3009837.3009897}
  {\path{doi:10.1145/3009837.3009897}}.

\bibitem[Mat13]{materzok-axiomatizing}
Marek Materzok.
\newblock Axiomatizing subtyped delimited continuations.
\newblock In {\em Computer Science Logic 2013}, CSL 2013. Schloss
  Dagstuhl-Leibniz-Zentrum fuer Informatik, 2013.

\bibitem[MB11]{materzok-subtyping}
Marek Materzok and Dariusz Biernacki.
\newblock Subtyping delimited continuations.
\newblock In {\em Proceedings of the 16th ACM SIGPLAN International Conference
  on Functional Programming}, ICFP '11, pages 81--93, New York, NY, USA, 2011.
  ACM.
\newblock \href {https://doi.org/10.1145/2034773.2034786}
  {\path{doi:10.1145/2034773.2034786}}.

\bibitem[Mil78]{milner-polymorphism}
Robin Milner.
\newblock A theory of type polymorphism in programming.
\newblock {\em Journal of computer and system sciences}, 17(3):348--375, 1978.

\bibitem[Nie01]{nielsen-selective}
Lasse~R Nielsen.
\newblock A selective {CPS} transformation.
\newblock {\em Electronic Notes in Theoretical Computer Science}, 45:311--331,
  2001.

\bibitem[Nor07]{norell-phd}
Ulf Norell.
\newblock {\em Towards a practical programming language based on dependent type
  theory}.
\newblock PhD thesis, Chalmers University of Technology, 2007.

\bibitem[Plo75]{plotkin-cps}
Gordon~D. Plotkin.
\newblock Call-by-name, call-by-value and the $\lambda$-calculus.
\newblock {\em Theoretical computer science}, 1(2):125--159, 1975.

\bibitem[PP09]{plotkin-handler}
Gordon Plotkin and Matija Pretnar.
\newblock Handlers of algebraic effects.
\newblock In {\em European Symposium on Programming}, ESOP '09, pages 80--94.
  Springer, 2009.

\bibitem[PPS19]{pirog-typedeq}
Maciej Pir{\'o}g, Piotr Polesiuk, and Filip Sieczkowski.
\newblock Typed equivalence of effect handlers and delimited control.
\newblock In {\em 4th International Conference on Formal Structures for
  Computation and Deduction (FSCD 2019)}. Schloss Dagstuhl-Leibniz-Zentrum fuer
  Informatik, 2019.

\bibitem[RMO09]{rompf-selective}
Tiark Rompf, Ingo Maier, and Martin Odersky.
\newblock Implementing first-class polymorphic delimited continuations by a
  type-directed selective {CPS}-transform.
\newblock In {\em Proceedings of the 14th ACM SIGPLAN International Conference
  on Functional Programming}, ICFP '09, pages 317--328, New York, NY, USA,
  2009. ACM.
\newblock \href {https://doi.org/10.1145/1596550.1596596}
  {\path{doi:10.1145/1596550.1596596}}.

\bibitem[SF93]{sabry-reasoning}
Amr Sabry and Matthias Felleisen.
\newblock Reasoning about programs in continuation-passing style.
\newblock {\em Lisp and symbolic computation}, 6(3):289--360, 1993.

\bibitem[Sha07]{shan-simulation}
Chung-chieh Shan.
\newblock A static simulation of dynamic delimited control.
\newblock {\em Higher-Order and Symbolic Computation}, 20(4):371--401, 2007.

\bibitem[SW97]{sabry-reflection}
Amr Sabry and Philip Wadler.
\newblock A reflection on call-by-value.
\newblock {\em ACM transactions on programming languages and systems (TOPLAS)},
  19(6):916--941, 1997.

\bibitem[WF94]{wright-syntactic}
Andrew~K. Wright and Matthias Felleisen.
\newblock A syntactic approach to type soundness.
\newblock {\em Information and computation}, 115(1):38--94, 1994.

\bibitem[WZD{\etalchar{+}}19]{wang-demystifying}
Fei Wang, Daniel Zheng, James Decker, Xilun Wu, Gr{\'e}gory~M Essertel, and
  Tiark Rompf.
\newblock Demystifying differentiable programming: Shift/reset the penultimate
  backpropagator.
\newblock {\em Proceedings of the ACM on Programming Languages}, 3(ICFP):1--31,
  2019.

\bibitem[XBH{\etalchar{+}}20]{xie-evidently}
Ningning Xie, Jonathan~Immanuel Brachth{\"a}user, Daniel Hillerstr{\"o}m,
  Philipp Schuster, and Daan Leijen.
\newblock Effect handlers, evidently.
\newblock {\em Proceedings of the ACM on Programming Languages}, 4(ICFP):1--29,
  2020.
\newblock \href {https://doi.org/10.1145/3408981} {\path{doi:10.1145/3408981}}.

\end{thebibliography}

\clearpage

\appendix

\section{Proof of Progress}
\label{app:progress}

In this appendix, we provide the detailed proof of the progress theorem 
for \lambdaF (Theorem 2).
The progress property of the fine-grained type system can be proved 
in a similar way.

\begin{lem}[Canonical Forms]
\label{lem:canonical}
If $\judge{\mtenv}{v}{(\arrowty{\tau_1}{\tau_2}{\mualpha}{\alpha}{\mubeta}{\beta})}
          {\mugamma}{\gamma}{\mugamma}{\gamma}$,
then $v$ is an abstraction $\abs{x}{e}$.
\end{lem}

\begin{proof}
By induction on the typing derivation.

\begin{case}[\textsc{Int}]
This case is impossible, since an integer cannot have an arrow type.
\end{case}

\begin{case}[\textsc{Var}]
This case is impossible, since a variable cannot be well-typed under
an empty environment.
\end{case}

\begin{case}[\textsc{Abs}]
This case is trivial, since the subject is an abstraction.
\end{case}

\begin{case}[Other]
Other cases are impossible, since the subject is not a value.
\end{case}
\par \vspace{-1.5\baselineskip}
\qedhere
\end{proof}

\begin{thm}[Progress]
If $\judge{\mtenv}{e}{\tau}{\mualpha}{\alpha}{\mubeta}{\beta}$,
then either (i) $e$ is a value, (ii) $e$ reduces to $\epr$, or
(iii) $e$ is a stuck term of the form $F[\control{k}{\epr}]$.
\end{thm}

\begin{proof}
By induction on the typing derivation.

\begin{case}[\textsc{Int}]
This case is trivial, since an integer is a value.
\end{case}

\begin{case}[\textsc{Var}]
This case is impossible, since no variable can be well-typed
under an empty environment.
\end{case}

\begin{case}[\textsc{Abs}]
This case is trivial, since an abstraction is a value.
\end{case}

\begin{case}[\textsc{App}]
By the induction hypothesis, we know that one of the following
holds for $e_1$.

\begin{enumerate}
\item $e_1$ is a value $v_1$
\item $e_1$ reduces to $\eonepr$
\item $e_1$ is a stuck term of the form $F[\control{k}{\eonepr}]$
\end{enumerate}

\noindent Similarly, we know that one of the following holds for $e_2$.

\begin{enumerate}
\setcounter{enumi}{3}
\item $e_2$ is a value $v_2$
\item $e_2$ reduces to $\etwopr$
\item $e_2$ is a stuck term of the form $F[\control{k}{\etwopr}]$
\end{enumerate}

Suppose 1 holds.
By Lemma~\ref{lem:canonical}, $e_1$ must be an abstraction $\abs{x}{\eonepr}$.
If 4 holds, the application is a $\beta$-redex.
If 5 holds, the application reduces to $\app{e_1}{\etwopr}$.
If 6 holds, the application is a stuck term $\Fpr[\control{k}{\eonepr}]$,
where $\Fpr$ = $\app{e_1}{F}$.

Next, suppose 2 holds.
In this case, the application reduces to $\app{\eonepr}{e_2}$.

Lastly, suppose 3 holds.
In this case, the application is a stuck term $\Fpr[\control{k}{\eonepr}]$,
where $\Fpr$ = $\app{F}{e_2}$.
\end{case}

\begin{case}[\textsc{Control}]
This case is trivial, since a \controltt expression is a stuck term.
\end{case}

\begin{case}[\textsc{Prompt}]
By the induction hypothesis, we know that one of the following holds
for $e$.

\begin{enumerate}
\item $e$ is a value $v$
\item $e$ reduces to $\epr$
\item $e$ is a stuck term of the form $F[\control{k}{\epr}]$
\end{enumerate}

Suppose 1 holds.
In this case, the \prompttt expression reduces to $v$.

Next, suppose 2 holds.
In this case, the \prompttt expression reduces to $\prompt{\epr}$.

Lastly, suppose 3 holds.
In this case, the \prompttt expression reduces to\\
$\prompt{\subst{e}{k}{\abs{x}{F[x]}}}$.
\end{case}
\par \vspace{-1.5\baselineskip}
\qedhere
\end{proof}

\end{document}